\documentclass[11pt]{article}
\usepackage{jheppub}

\usepackage{amsmath}
\usepackage{amssymb}
\usepackage{verbatim}

\usepackage{multirow}

\usepackage[T1,OT1]{fontenc}

\usepackage{graphicx}

\allowdisplaybreaks[4]

\usepackage[hang]{footmisc}
\renewcommand{\d}{\mathrm{d}}
\newcommand{\e}{\mathrm{e}}
\newcommand{\w}{\wedge}

\newcommand{\nl}{\notag \\ &\quad\,}

\title{de Sitter-eating O-planes in supercritical string theory}

\author{Daniel Junghans}

\affiliation{Institute for Theoretical Physics, TU Wien, Wiedner Hauptstra\ss e 8-10/136, A-1040 Vienna, Austria}

\emailAdd{daniel.junghans @ tuwien.ac.at}

\notoc

\abstract{It has been proposed that flux compactifications of supercritical string theories (i.e., with spacetime dimension $D>10$) have dS vacua, with large $D$ acting as a control parameter for corrections to the classical spacetime effective action.
In this paper, we provide a detailed analysis of the self-consistency of such models, focussing on $\alpha^\prime$ and backreaction corrections.
We first show that all supercritical AdS, Minkowski and dS vacua in this setting have $\gtrsim \mathcal{O}(1)$ curvature and/or field strengths in the string frame.
This may
be in tension with suppressing $\alpha^\prime$ corrections
unless the coefficients of the higher-derivative terms have a sufficiently strong large-$D$ suppression.
We then argue that an additional and more severe problem arises in the dS case due to the backreaction of O-planes.
In particular, we argue using a combination of geometric bounds and string-theory constraints that the O-plane backreaction is
large in supercritical dS models. This implies that a large part of the naive classical geometry is eaten up by singular holes and thus indicates a breakdown of the classical description.
Our finding resonates with several other recent results suggesting that string theory does not admit dS vacua in regimes where string and backreaction corrections are under control.
As byproducts of our analysis, we derive a number of technical results that are useful beyond the specific applications in this paper. In particular, we compute the leading backreaction corrections to the smeared solution in a general flux compactification from $D$ to $d$ dimensions for an arbitrary distribution of O-planes and D-branes. We further argue for a general estimate for Green's functions on compact manifolds (and therefore for the backreaction corrections) in terms of their diameter, volume and dimension.
}

\begin{document}

\numberwithin{equation}{section}

\maketitle

\newpage

\section{Introduction}

One of the most basic yet mysterious properties of our universe is
its accelerated expansion. If string theory is to successfully describe
nature,
it should certainly admit solutions reproducing this behavior. The simplest possibility would be a compactification to dS space but none of the proposed scenarios in string theory have been developed into fully explicit models to date.
Even worse, a number of recent works raised serious doubts about the self-consistency of the
approximation schemes underlying the most popular scenarios, i.e., the KKLT scenario \cite{Kachru:2003aw}, the LARGE-volume scenario \cite{Balasubramanian:2005zx} and the classical dS scenario \cite{Silverstein:2007ac, Hertzberg:2007wc}.
A recurring theme in all three scenarios
is that it is very difficult, if not impossible, to simultaneously suppress all relevant $\alpha^\prime$ and backreaction corrections whenever the vacuum energy is naively positive \cite{Carta:2019rhx, Gao:2020xqh, Blumenhagen:2022dbo, Junghans:2022exo, Gao:2022fdi, Junghans:2022kxg, Hebecker:2022zme, Schreyer:2022len, ValeixoBento:2023nbv, Junghans:2018gdb, Banlaki:2018ayh, Cribiori:2019clo, Andriot:2019wrs}.\footnote{Further recent debates about KKLT concern, e.g., the consistency of the pre-uplift AdS vacua \cite{Demirtas:2021nlu, Lust:2022lfc} and possible instabilities of the anti-brane uplift due to goldstino condensation \cite{DallAgata:2022abm, Kallosh:2022fsc}.} This leads to a loss of control and consequently a breakdown of the effective field theories in which the putative dS vacua arise. By contrast, in \emph{Anti-}dS vacua such as DGKT \cite{DeWolfe:2005uu, Camara:2005dc} a parametric control over both string and backreaction corrections appears to be possible \cite{Junghans:2020acz, Marchesano:2020qvg, Cribiori:2021djm}.

A crucial question is whether the reported difficulties with dS are a lamppost effect
or whether there is something fundamentally problematic about dS space, as proposed in various works \cite{Danielsson:2018ztv, Obied:2018sgi, Ooguri:2018wrx, Dvali:2017eba, Dvali:2018jhn}.
To make progress, it is important to get off the beaten track of the standard scenarios
and explore as many regimes of string theory as possible.
We stress that establishing \emph{any} clear and explicit construction of a dS vacuum
would be a very welcome proof of principle and a major conceptual breakthrough. On the other hand, purely phenomenological constraints such as a realistic value of the cosmological constant or a four-dimensional spacetime are not a priority from this point of view.

In this paper, we will study a rarely explored
corner of the string landscape, namely supercritical string theories in $D>10$ dimensions \cite{Myers:1987fv, Antoniadis:1988aa, deAlwis:1988pr}. We will consider compactifications of these theories to $d$ dimensions with $4\le d\le D$ including fluxes and localized sources (D-branes and O-planes).
It has been argued that such models can be described in terms of a $D$-dimensional spacetime effective action (or a dimensional reduction thereof to $d$ dimensions), in analogy to the standard approach in the critical string theories \cite{deAlwis:1988pr, Polchinski:1998rq, Polchinski:1998rr, Silverstein:2001xn}.\footnote{See also 
\cite{Silverstein:2005qf, McGreevy:2006hk, Green:2007tr, Hellerman:2004zm, Hellerman:2004qa, Hellerman:2006ff, Hellerman:2007fc, Bonnefoy:2020uef, DeBiasio:2021yoe, Lust:2021mgj} for discussions of dualities relating supercritical and critical theories and \cite{Bonnefoy:2020uef} for a discussion of swampland conjectures at large $D$.}

Supercritical flux compactifications differ in a number of interesting ways from their more familiar critical cousins.
In particular, a distinctive feature
is the appearance of an additional term in the scalar potential that generates a tadpole for the dilaton.
The moduli dependence of the potential thus differs from the one in the critical theories, leading to new possibilities for moduli stabilization.
In fact, it was argued in \cite{Silverstein:2001xn, Maloney:2002rr, Dodelson:2013iba, Harribey:2018xvs} that simple dS solutions arise in this setting.

Another interesting point is that the large dimension $D$ can serve as an expansion parameter, which may be useful to control corrections to the classical effective action. Indeed, there are general field-theoretic arguments suggesting that
(quantum) gravity simplifies
at large $D$
\cite{Strominger:1981jg, Emparan:2013moa}.
In the context of string theory, it was proposed in \cite{Silverstein:2001xn, Maloney:2002rr, Harribey:2018xvs, DeLuca:2021pej, Flauger:2022hie} that string corrections might be parametrically controlled at large $D$.
Furthermore, in orientifolds, the number of O-planes becomes exponentially large in the large-$D$ regime (e.g., $N_{\text{O}p}=2^{D-p-1}$ on a torus).
One might thus hope \cite{Silverstein:2001xn, Maloney:2002rr, Dodelson:2013iba} that generating large parameters in the scalar potential is easier than in the critical string theories where the possibility to do so is often limited by rather small tadpoles. Recall that a potential with large parameters is crucial in order to avoid the Dine-Seiberg problem and allow vacua in regimes of perturbative control \cite{Dine:1985he, Denef:2008wq}.

On the other hand, it is natural to suspect that the aforementioned control problems in the dS scenarios of the critical string theories point towards a more general pattern in string theory. This suggests that supercritical dS vacua may fall victim to the same problems, in particular large string or backreaction corrections that invalidate the naive effective field theories in which these vacua arise. The purpose of the present work is to perform a detailed study of this hypothesis, where we focus on parametric control at large $D$ in models based on the $D$-dimensional spacetime effective action, as proposed in the works mentioned above.
We will indeed argue that some of the consistency conditions one should impose for the classical effective action are violated.
While some of our results apply to all kinds of vacua (i.e., AdS, Minkowski or dS), the problems we find are more severe in the dS case, which is related to the fact that dS vacua require O-planes.

In the first part of the paper, we attempt to quantify the magnitude of $\alpha^\prime$ corrections, which must be small in order that computations based on the classical effective action are trustworthy.
In particular, we show that the string-frame curvature and/or some of the NSNS and RR energy densities are always large (in string units) for all supercritical AdS, Minkowski and dS vacua. This follows from the simple observation that the equations of motion contain terms $\sim (D-10)\mathcal{O}(1)$ due to the dilaton tadpole and can therefore only be satisfied if some of the other terms are of the same order. One may be tempted to interpret this as a breakdown of the $\alpha^\prime$ expansion and conclude that the vacua cannot be trusted at the classical level.
However, as pointed out in \cite{Harribey:2018xvs, DeLuca:2021pej, Flauger:2022hie}, large scalar curvature and energy densities may be ok at large $D$
if the (unknown) coefficients of the most dangerous higher-derivative corrections vanish or scale with $D$ in such a way that an additional suppression is provided.
Our result translates to upper bounds on these scalings, which would have to be satisfied in string theory in order that supercritical flux vacua can be reliable.
In some cases, we furthermore observe a tension with the consistency condition that massive string states should be heavier than the KK scale.

In the second part of our paper, we give an independent second argument for the breakdown of the effective action which
is less sensitive to
assumptions about higher-derivative terms. This argument applies only to dS vacua and is based on the observation that the backreaction of O-planes makes a classical solution unreliable if it is too large.
Note that O-planes are required for dS (in the regime where $\alpha^\prime$ corrections are subleading) according to a well known no-go theorem \cite{Gibbons:1984kp, deWit:1986xg, Maldacena:2000mw}, which we generalize to arbitrary $D$.
In order to estimate the backreaction corrections, we show following \cite{Junghans:2020acz} that they are determined by the Green's function of the Laplacian on the transverse space of the corresponding source. We then provide an extensive discussion of Green's functions and give evidence for a general estimate in terms of the diameter, the volume and the dimension of the transverse space. These results apply to localized sources in critical and supercritical string theories and should be useful for various applications in future works.

Using our general estimate for the backreaction, we then argue that it is large in a wide variety of supercritical dS models.
This is a problem for the consistency of these models since O-planes are known to create singular holes in their vicinity where string effects blow up and the classical approximation is no longer valid.
In limits of small backreaction, these holes shrink until they do not significantly affect the classical solution in most of the spacetime.
For example, this can explicitly be shown in DGKT \cite{Junghans:2020acz, Marchesano:2020qvg}. However, our results indicate that regimes of small backreaction are not possible in supercritical dS models where instead the singular holes eat up a large fraction of the classical geometry,
and consequently the dS vacua cannot be trusted.

Interestingly, the path leading to
this conclusion can be rather different depending on the
model, and it
generally involves a non-trivial
interplay between all kinds of
constraints from both geometry and physics.
For example, we employ rigorous theorems on diameters, volumes, Laplacian eigenvalues and Betti numbers
as well as constraints from string theory such as flux quantization, tadpole cancellation, anomaly cancellation and an absence of tachyons.\footnote{See also \cite{Gautason:2015tig, DeLuca:2021mcj, DeLuca:2021ojx, DeLuca:2022wfq, DeLuca:2023kjj, Collins:2022nux} for other interesting applications of diameter and eigenvalue bounds in the context of string theory.}
Our results thus exemplify
the familiar fact that string theory often finds
highly non-trivial ways of obstructing seemingly consistent dS scenarios as one digs into the details.

Another challenging aspect of our analysis is that we have to
carefully keep track of $D$-dependent factors in every step of our arguments.
Correctly accounting for such factors is crucial at large $D$ since they could parametrically change the conclusions if missed.
For example, the external dimension $d$, the dimensions of the localized sources or the ranks of the RR fluxes
can all scale non-trivially with $D$ and thus affect the magnitude of various terms in the equations of motion. Another example is that,
even on simple spaces like spheres, one has to distinguish the different length scales set by the diameter, the curvature and the volume since, at large $D$, they all differ parametrically from one another.
We pay particular attention to such subtleties throughout the paper
and formulate our arguments without assuming specific relations between these (and other) geometric quantities.

Our focus in this paper is on geometric models, i.e., models where the $(D-d)$-dimensional compactification space is a manifold or orientifold with a metric whose dynamics is governed by the Einstein equations. We will occasionally also comment on the dS models of \cite{Silverstein:2001xn, Maloney:2002rr} where one compactifies on an asymmetric orbifold \cite{Narain:1986qm} so that the internal space does not admit a geometric interpretation. Interestingly, these models satisfy the same bounds and inequalities that we find for the geometric models. In particular, they can be shown to have large RR energy densities. As discussed above, this may be in tension 
with controlling higher-derivative terms depending on the large-$D$ scaling of the corresponding coefficients.
The models further have, at least formally, a large O-plane backreaction just like their geometric cousins. Although it is less obvious what large backreaction means without an internal geometry, we find it plausible to again interpret it as a control problem, as we will discuss in more detail in the main text.

This paper is organized as follows.
In Section \ref{sec:setup}, we review some basic facts about supercritical string theories that will be relevant for our analysis. We also derive a generalization of the no-go theorem of \cite{Gibbons:1984kp, deWit:1986xg, Maldacena:2000mw} valid for arbitrary $D$ and compute the scalar potential with respect to the volume and dilaton moduli. In Section \ref{sec:string}, we show that classical supercritical AdS, Minkowski and dS vacua necessarily have large string-frame curvature or large NSNS/RR energy densities. We then explain why this may indicate a breakdown of the $\alpha^\prime$ expansion and review a conjectured loophole to this conclusion at large $D$. We also discuss a related tension with the requirement of avoiding light string states below the KK scale.
In Section \ref{sec:backreaction}, we compute backreaction corrections of localized sources in critical and supercritical string theories.
We also propose an estimate for Green's functions on compact manifolds in terms of their diameter, volume and dimension.
We further argue that supercritical dS models have a large O-plane backreaction and that this indicates a breakdown of the classical effective action.
In Section \ref{sec:discussion}, we discuss how our results relate to other recent works on dS scenarios in string theory. We conclude in Section \ref{sec:concl} with a few remarks about future research. In App.~\ref{app:mink}, we describe Minkowski solutions of the classical equations of motion which, in contrast to the critical type II string theories, do not require any O-planes. In App.~\ref{app:caveat}, we elaborate on a subtlety related to the discussion of singular holes and their stringy resolution in Section \ref{sec:b1}.

\section{Setup}
\label{sec:setup}

\subsection{Effective action and equations of motion}

A well known consistency condition for the quantization of the superstring is that the total central charge of the worldsheet CFT vanishes. In flat space, this leads to the famous requirement of a critical spacetime dimension $D=10$. However, it has been argued that $D\neq 10$ is possible when the strings propagate in a non-trivial background since then curvature and field gradients contribute to the central charge and modify the condition on $D$ \cite{Myers:1987fv, Antoniadis:1988aa, deAlwis:1988pr, Polchinski:1998rq, Polchinski:1998rr}.

In this paper, we consider supercritical ($D>10$) generalizations of the type II string theories following \cite{Silverstein:2001xn, Maloney:2002rr, Dodelson:2013iba}. These theories are consistent for $D=10 \text{ (mod } 16\text{)}$ \cite{Dodelson:2013iba}. The $D$-dimensional spacetime effective action in string frame is \cite{deAlwis:1988pr, Polchinski:1998rq, Polchinski:1998rr, Silverstein:2001xn}
\begin{align}
S &= \frac{1}{2\kappa_D^2} \int \d^D x \sqrt{-g_D}\, \e^{-2\phi} \left(\mathcal{R}_D - \frac{D-10}{\alpha^\prime} +4(\partial\phi)^2  -\frac{1}{2}|H_3|^2 - \frac{1}{4}\sum_q \e^{2\phi}|F_q|^2 \right), \label{action}
\end{align}
where we choose $2\kappa_D^2 = \frac{(2\pi\sqrt{\alpha^\prime})^{D-2}}{2\pi}$.\footnote{The results of Sections \ref{sec:mn}, \ref{sec:stab} and \ref{sec:string} also apply to the bosonic string upon substituting $D-10\to \frac{2}{3}(D-26)$ and setting the localized sources and RR terms to zero.}
The index $q$ runs from 0 to $D$ over either the even or odd numbers as in the (democratic formulation of) the critical theories \cite{Dodelson:2013iba}.\footnote{We use the duality relation $\star_D F_q = (-1)^{q(q-1)/2} F_{D-q}$ and assume following \cite{Dodelson:2013iba} that $F_q =\d C_{q-1}-H_3\w C_{q-3}$ (away from localized sources and up to Romans-mass terms) as in the critical theories.}
Fluxes are quantized as \cite{Maloney:2002rr}
\begin{equation}
\frac{1}{2\kappa_D}\int F_q = \sqrt{\pi} (2\pi\sqrt{\alpha^\prime})^{q-\frac{D}{2}} \mathbb{Z}
\end{equation}
and analogously for $H_3$.

We will also include D$p$-branes and O$p$-planes wrapping submanifolds $\Sigma_i$
of the $D$-dimensional spacetime. Their action is
\begin{equation}
S_\text{loc} =  - T_\text{O$p$/D$p$} \left(\int_{\Sigma_i} \d^{p+1} x \sqrt{-g_{p+1}}\, \e^{-\phi}- \int_{\Sigma_i} C_{p+1}\right) \label{sloc}
\end{equation}
with (downstairs) tension/charge \cite{Silverstein:2001xn}\footnote{We divided the formula in \cite{Silverstein:2001xn} by $4$ in order that it matches for $D=10$ with the standard result in \cite{Polchinski:1998rr}.
The difference is not important in this paper since our arguments will not require specific values for $T_\text{O$p$/D$p$}$. We further assume that the RR charge, which was not explicitly computed in \cite{Silverstein:2001xn}, is determined by the tension as in the critical theories.
}
\begin{equation}
T_\text{O$p$} = - 2^{p-D/2}T_\text{D$p$}, \qquad T_\text{D$p$} = \frac{2^{5/2-D/4} 2\pi}{(2\pi\sqrt{\alpha^\prime})^{p+1}}. \label{tp}
\end{equation}
As in the critical type II string theories, $p$ runs either over the even or odd numbers.
When considering anti-branes or anti-O-planes, the sign of the $C_{p+1}$ term in \eqref{sloc} has to be flipped. Note that we omitted couplings to the $B_2$ field in the (anti-)brane action, as they play no role in the following. We will furthermore set $2\pi\sqrt{\alpha^\prime}=1$ in all following expressions.

We consider compactifications of the $D$-dimensional theories down to $4\le d \le D$ dimensions.\footnote{Most of our results carry over to the case $d=3$ as well with the qualification that then the $H_3$ flux is allowed to fill the $d$-dimensional spacetime, which yields an additional term in the equations of motion.} Since we are interested in vacuum solutions, we assume that maximal symmetry is preserved in $d$ dimensions. This implies that the D$p$-branes/O$p$-planes must fill the $d$-dimensional spacetime and a $(p+1-d)$-dimensional submanifold of the compact internal space. Furthermore, derivatives of the various fields with respect to the $d$ external dimensions must vanish on-shell, and each component of $H_3$ and $F_q$ must have either zero or $d$ legs in the $d$-dimensional spacetime. In the following, we choose to write all RR field strengths with external legs in terms of their internal Poincar\'e duals. The index $q$ then runs from 0 to $D-d$ in these conventions.

The most general ansatz for the metric is
\begin{equation}
\d s_D^2 = \e^{2A(x^m)} \d s_d^2 + g_{mn} \d x^m \d x^n, \label{metric}
\end{equation}
where $\d s_d^2$ is the $d$-dimensional AdS, Minkowski or dS metric and $\e^{2A}$ is a function of the internal coordinates called the warp factor.

The dilaton equation and the (trace-reversed) external and internal Einstein equations are
\begin{align}
0 &= -8 \e^{\phi} \nabla^2 \e^{-\phi} -4d\e^{-A}\nabla^2 \e^A  -2d(d-1)( \partial A)^2 + 8d ( \partial A \cdot \partial \phi) + 2\e^{-2A} \mathcal{ R}_d + 2\mathcal{ R}_{D-d} \nl - 8\pi^2 (D-10) - |H_3|^2 - \sum_i \frac{T_{p_i}}{2\pi} \e^\phi \delta(\Sigma_i), \label{eom1} \\
0 &=  \frac{2d}{D-2}\e^{\phi}\nabla^2\e^{-\phi} +d\e^{-A}\nabla^2\e^A +\frac{2d}{D-2}( \partial \phi)^2+d(d-1)( \partial A)^2 -\frac{2d(D+d-2)}{D-2}( \partial A \cdot \partial \phi) \nl - \e^{-2A} \mathcal{R}_d + \frac{d}{D-2} 4\pi^2 (D-10) - \frac{d}{D-2} |H_3|^2 - \sum_{q} \frac{(q-1)d}{2(D-2)}\e^{2\phi} |F_q|^2 \nl - \sum_i \frac{T_{p_i}}{2\pi} \frac{(D-p_i-3)d}{2(D-2)} \e^\phi \delta(\Sigma_i), \label{eom2} \\
0 &= 2\e^{\phi} \nabla_m\partial_n \e^{-\phi} + d\e^{-A}\nabla_m\partial_n \e^A + \frac{2}{D-2} g_{mn}\e^{\phi} \nabla^2\e^{-\phi} + \frac{2}{D-2} g_{mn} (\partial\phi)^2 \nl -\frac{2d}{D-2} g_{mn}(\partial A\cdot\partial\phi) - 2(\partial_m\phi)(\partial_n\phi)  -\mathcal{R}_{mn}
 + g_{mn}\frac{4\pi^2}{D-2} (D-10) + \frac{1}{2}|H_3|^2_{mn} \nl - \frac{1}{D-2}g_{mn}|H_3|^2 + \frac{1}{2} \e^{2\phi} \sum_{q} \left( |F_q|^2_{mn}- \frac{q-1}{D-2}g_{mn}|F_q|^2 \right) \nl - \frac{1}{2}\sum_i \frac{T_{p_i}}{2\pi} \left(\Pi^{(i)}_{mn}-\frac{p_i+1}{D-2}g_{mn}\right) \e^\phi \delta(\Sigma_i). \label{eom3}
\end{align}
Here $T_{p_i}\in\{ T_{\text{O}p_i}, T_{\text{D}p_i}\}$ depending on the source and $\Pi^{(i)}_{mn}$ is defined such that $\Pi^{(i)}_{mn}=g_{mn}$ for indices parallel to the source and $\Pi^{(i)}_{mn}=0$ for transverse indices. The covariant derivative $\nabla_m$ and the Laplacian $\nabla^2$ are adapted to the $(D-d)$-dimensional metric $g_{mn}$.
We furthermore denote by $\mathcal{R}_d$ and $\mathcal{R}_{D-d}$ the external and internal parts of the Ricci scalar with respect to the unwarped metric.
The delta distributions are due to the localized sources and defined such that
\begin{equation}
\int \d^D x \sqrt{-g_D}\, f \delta(\Sigma_i) = \int_{\Sigma_i} \d^{p_i+1} x \sqrt{-g_{p_i+1}}\, f|_{\Sigma_i}, \label{delta}
\end{equation}
where $f$ is some function of the internal coordinates and $f|_{\Sigma_i}$ is its restriction to $\Sigma_i$. The above definition implies that $\delta(\Sigma_i) = \frac{\delta(\vec x)}{\sqrt{g_{D-p_i-1}}}$ in local coordinates, including an inverse determinant of the metric of the transverse space.
In the smeared approximation, one replaces
\begin{equation}
\delta(\Sigma_i) \to \frac{1}{V_i} \label{smear}
\end{equation}
in the equations of motion, where $V_i$ is the volume of the transverse space. This corresponds to only taking into account the zero mode of $\delta(\Sigma_i)$ and neglecting the effect of the higher modes. The smeared solution is thus reliable at distances from the source which are longer than the backreaction length of the higher modes, as we will discuss in more detail in Section \ref{sec:b1}.

We finally state the equations of motion and Bianchi identities for the RR and NSNS form fields:
\begin{align}
&\d \left(\star_D F_q\right) = - H_3 \w \star_D F_{q+2}, \qquad \d F_q = H_3 \w F_{q-2} - (-1)^{q(q+1)/2} \sum_{i}\frac{Q_{p_i}}{2\pi} \delta_{D-p_i-1}, \label{eom4} \\
&\d \left(\e^{-2\phi}\star_D H_3\right) = -\sum_q \star_D F_q \w F_{q-2}, \qquad \d H_3 =0, \label{eom5}
\end{align}
where $Q_{p_i}=\pm T_{p_i}$ (the lower sign is for anti-branes and anti-O-planes) and $\delta_{D-p_i-1} = \delta(\Sigma_i) \text{dvol}_{D-p_i-1}$. The sum in the second equation in \eqref{eom4} runs over all $i$ with $p_i=D-q-2$.
The Hodge star $\star_D$ is defined with respect to the full $D$-dimensional metric \eqref{metric} including the warp factor.

\subsection{No-go theorem against dS without O-planes}
\label{sec:mn}

It is well known that classical dS vacua of the critical type II string theories require O-planes \cite{Gibbons:1984kp, deWit:1986xg, Maldacena:2000mw}. We now prove a straightforward generalization of this no-go theorem which is valid for arbitrary $D$. Indeed, combining \eqref{eom1}--\eqref{eom3}, we find
\begin{equation}
\e^{-2A}\mathcal{R}_d-\e^{2\phi-dA}\nabla_m \left(\e^{-2\phi} \partial^m \e^{dA}\right) = - \frac{d}{4} \sum_{q} \e^{2\phi} |F_q|^2 - \frac{d}{4} \sum_i \frac{T_{p_i}}{2\pi} \e^\phi \delta(\Sigma_i),
\end{equation}
where $\mathcal{ R}_d>0$ in a dS vacuum.
Multiplying both sides by $\e^{dA-2\phi}$ and integrating over the compact space, we find
\begin{equation}
\mathcal{R}_d = - \frac{d}{4} \sum_{q} \langle\e^{2\phi} |F_q|^2\rangle - \frac{d}{4} \sum_i \frac{T_{p_i}}{2\pi} \langle \e^\phi \delta(\Sigma_i)\rangle, \label{mn}
\end{equation}
where we introduced the notation
\begin{equation}
\langle X \rangle = \frac{\int \d^{D-d}x \sqrt{g_{D-d}}\,\e^{dA-2\phi} X}{\int \d^{D-d}x \sqrt{g_{D-d}}\,\e^{(d-2)A-2\phi}}. \label{avr}
\end{equation}
Since the RR terms on the right-hand side of \eqref{mn} are negative, we have $\mathcal{R}_d\le 0$
unless there are sources with $T_{p_i}=T_{\text{O}p_i}<0$. A necessary condition for a dS vacuum is thus the presence of O-planes.
More precisely, an O-plane filling the full $D$-dimensional spacetime would not be sufficient to circumvent the no-go as its RR tadpole would have to be cancelled by D-branes,
which would at the same time also cancel the source term in \eqref{mn}. We therefore require O-planes with codimension $\ge 1$ (if the tadpole is cancelled by anti-O-planes) or $\ge 3$ (if the tadpole is cancelled by fluxes) to get dS.

An interesting subtlety occurs for the Minkowski case $\mathcal{R}_d= 0$. Unlike in the critical type II theories \cite{Giddings:2001yu}, the equations of motion admit Minkowski solutions with $H_3$ flux but \emph{without} RR flux and O-planes.
These solutions are described in more detail in App.~\ref{app:mink}.

We finally note that, aside from O-planes, a second possibility to circumvent the dS no-go is to go beyond the classical effective action \eqref{action}, \eqref{sloc} and include corrections such as higher-derivative terms. We will not study this possibility in this paper since, even at next-to-leading order in the $\alpha^\prime$ expansion, the knowledge about the corrections is incomplete (even more in the supercritical case). It would therefore be difficult to verify the consistency of solutions which rely on higher-derivative terms.\footnote{An alternative scenario proposed in \cite{Harribey:2018xvs} is to use one-loop corrections in the two-derivative effective action of the supercritical bosonic string. The dS solutions found in \cite{Harribey:2018xvs} are unstable due to the usual tachyon of the bosonic string and an additional instability of the dilaton. Furthermore, the coefficients of the relevant loop corrections are not known for general $D$ and were assumed to be free parameters in \cite{Harribey:2018xvs}. Finally, the solutions have large string-frame curvature so that the problems discussed below in Section \ref{sec:string} may apply. More work is therefore required to see whether controlled (meta-)stable dS vacua can arise from loop corrections at the two-derivative level (in those string theories where such corrections exist).
}

\subsection{Scalar potential and stability}
\label{sec:stab}

In order to study the stability of putative vacua, it is useful to derive the scalar potential of the $d$-dimensional effective field theory. For simplicity, we do this in the regime where the smeared approximation is accurate on most of the compact space so that backreaction corrections do not significantly affect the scalar potential. We will argue in Section \ref{sec:backreaction} that the existence of such a regime is actually a crucial consistency condition for solutions of the classical equations of motion with O-planes.
Since we will only impose stability conditions for solutions with O-planes in this paper, our assumption of a smeared regime is therefore justified.

In the smeared limit, the warp factor and the dilaton are constant, which significantly simplifies the various expressions. For our purpose, it will furthermore be sufficient to only keep track of two moduli in the scalar potential, i.e., the dilaton $\tau\equiv \e^{-\phi}$ and the volume $\rho$ of the $(D-d)$-dimensional compact space.
In general, the scalar potential of course depends on many more moduli but we choose to leave the dependence on these moduli implicit.

To derive the potential, we perform a dimensional reduction of the action \eqref{action}, \eqref{sloc} down to $d$ dimensions.\footnote{Note that a naive dimensional reduction of the democratic (pseudo-)action would lead to a wrong scalar potential with vanishing RR terms, as is familiar, e.g., from the self-dual $F_5$ in type IIB string theory. The correct procedure is to either use a formulation of the action which is not subject to a duality constraint or read off the scalar potential from the external Einstein equations (see, e.g., \cite{DeWolfe:2002nn, Giddings:2005ff} for a discussion in the type IIB context).}
This yields
\begin{align}
S \supset 2\pi \int \d^d x \sqrt{-g_d}\, &\Big( \tau^2 \rho \mathcal{R}_d + \tau^2 \rho^{\frac{D-d-2}{D-d}} \mathcal{\hat R}_{D-d}
 - 4\pi^2 \tau^2 \rho\, (D-10) -\frac{1}{2}\tau^2 \rho^{\frac{D-d-6}{D-d}} |\hat H_3|^2 \nl - \frac{1}{2}\sum_q\rho^{\frac{D-d-2q}{D-d}} |\hat F_q|^2 - \sum_i \frac{T_{p_i}}{2\pi\hat V_i} \tau \rho^{\frac{p_i+1-d}{D-d}} \Big).
\end{align}
Here we extracted the volume dependence from the various terms by defining $g_{mn}= \rho^{\frac{2}{D-d}}\hat g_{mn}$ such that $\hat g$ has unit volume and $\rho = \int \d^{D-d}x\, \sqrt{g_{D-d}}$ denotes the volume modulus.

We now go to the $d$-dimensional Einstein frame by performing the further redefinition
\begin{equation}
g_{\mu\nu} =  \tau^{-\frac{4}{d-2}} \rho^{-\frac{2}{d-2}} \hat g_{\mu\nu}.
\end{equation}
We thus find
\begin{equation}
S \supset 2\pi \int \d^d x \sqrt{-\hat g_d} \left( \mathcal{\hat R}_d - V\right)
\end{equation}
with scalar potential
\begin{align}
V &= -\tau^{-\frac{4}{d-2}} \rho^{\frac{D-d-2}{D-d} - \frac{d}{d-2}} \mathcal{\hat R}_{D-d} + 4\pi^2 \tau^{-\frac{4}{d-2}} \rho^{-\frac{2}{d-2}} (D-10) + \frac{1}{2}\tau^{-\frac{4}{d-2}} \rho^{\frac{D-d-6}{D-d} - \frac{d}{d-2}} |\hat H_3|^2 \nl + \frac{1}{2}\sum_q\tau^{- \frac{2d}{d-2}} \rho^{\frac{D-d-2q}{D-d} -\frac{d}{d-2}} |\hat F_q|^2 + \sum_i \frac{T_{p_i}}{2\pi \hat V_i} \tau^{-\frac{d+2}{d-2}} \rho^{\frac{p_i+1-d}{D-d} -\frac{d}{d-2}}.
\end{align}
The $d$-dimensional traced Einstein equation and the two universal moduli equations are
\begin{equation}
\mathcal{\hat R}_d - \frac{d}{d-2}V = 0, \qquad \partial_\tau V=0, \qquad \partial_\rho V=0.
\end{equation}
It is straightforward to check that these agree with \eqref{eom1}--\eqref{eom3} after undoing the previous field redefinitions of the metric.

An advantage of the formulation in terms of a scalar potential is that the conditions for the stability of a solution become very simple. In particular, we will impose below the requirement that the dilaton is stabilized. In the dS/Minkowski case, we require $\partial_\tau^2 V >0$\footnote{
In AdS, $\partial_\tau^2 V<0$ is allowed as long as the Breitenlohner-Freedman bound \cite{Breitenlohner:1982bm, Breitenlohner:1982jf} is not violated. For dS, an analogous \emph{upper} bound on scalar masses was furthermore proposed in \cite{Strominger:2001pn}.}
and thus
\begin{align}
-\frac{d+2}{8}g_s\sum_i \frac{T_{p_i}}{2\pi V_i} &< \frac{d}{4}g_s^2\sum_q |F_q|^2 && \text{(dS/Minkowski)}, \label{stab}
\end{align}
where we used $\partial_\tau V=0$ to simplify the inequality. We also replaced $\tau$ by its vev $1/g_s$ and transformed back to the unhatted metric $g_{mn}$.

Recall that, according to \eqref{mn}, the net tension of the source terms must be negative (i.e., $\sum_i\frac{T_{p_i}}{V_i} < 0$) in dS/Minkowski vacua if RR terms are present. \eqref{stab} therefore implies
that the source terms in dS/Minkowski vacua cannot be parametrically larger than the RR terms, which we will repeatedly use below. For AdS with net-negative-tension sources, the same conclusion follows from \eqref{mn} alone.

\section{$\alpha^\prime$ corrections}
\label{sec:string}

In this section, we show that the string-frame curvature and/or the energy densities of the RR/NSNS fluxes are large in vacuum solutions of the supercritical string theories.
This applies to AdS, Minkowski and dS vacua.
We also discuss to what extent this result is in tension with suppressing string corrections to the classical effective action.

\subsection{Large curvature and energy densities}
\label{sec:string1}

Let us again focus on smeared solutions for simplicity. In solutions with localized sources (i.e., with varying dilaton and warp factor), analogous arguments can be made for averaged energy densities (such as in \eqref{mn}).

Using \eqref{eom1}, \eqref{eom2} and the trace of \eqref{eom3}, we find
\begin{align}
\mathcal{R}_{D-d} &= \frac{d+2}{4}g_s\sum_i \frac{T_{p_i}}{2\pi V_i}  + \frac{1}{2}|H_3|^2 + \frac{d}{4}g_s^2\sum_{q} |F_q|^2 + 4\pi^2 (D-10), \label{f1} \\
|H_3|^2 &= -\frac{1}{4}g_s \left(D-2\overline{p}-4\right) \sum_i\frac{T_{p_i}}{2\pi V_i} + g_s^2 \frac{D-2\overline{q}}{4} \sum_{q} |F_q|^2 + 4\pi^2 (D-10). \label{f2}
\end{align}
Here we defined an effective RR rank and an effective source dimension (similar to \cite{Dodelson:2013iba}) such that
\begin{equation}
\overline{q} \equiv \frac{\sum_q q |F_q|^2}{\sum_q |F_q|^2}, \qquad \overline{p}\equiv \frac{\sum_i p_i \frac{T_{p_i}}{V_i}}{\sum_i \frac{T_{p_i}}{V_i}}
\end{equation}
if $\sum_q |F_q|^2\neq 0$ and $\sum_i \frac{T_{p_i}}{V_i}\neq 0$, respectively (and $\overline{q}=0$, $\overline{p} \sum_i \frac{T_{p_i}}{V_i} \to \sum_i p_i \frac{T_{p_i}}{V_i}$ otherwise).
Note that $q_\text{min}\le \overline{q}\le q_\text{max}$, where $q_\text{min/max}$ denotes the smallest/largest $q$ with $F_q\neq 0$.
On the other hand, $\overline{p}$ does not satisfy an analogous inequality if sources with both signs of the tension are present. In particular, $\overline{p}$ can be larger than $p_\text{max}$ or smaller than $p_\text{min}$, where $p_\text{min/max}$ denotes the smallest/largest $p_i$ with $T_{p_i}\neq 0$.
Neither $\overline{p}$ nor $\overline{q}$ need to be integers in general.

Let us now analyze the constraints on the curvature/energy densities imposed by \eqref{f1} and \eqref{f2}.
We will ignore $\mathcal{O}(1)$ numerical factors and focus on the parametric scaling of the terms with $D$ in the regime where $D$ is large. We will also keep track of the dependence on $d$, $\overline{q}$ and $\overline{p}$, as each of these numbers may a priori scale non-trivially with $D$.
It is furthermore convenient to distinguish three different cases depending on the sign and magnitude of the source terms. The latter will be denoted as small (large) if $g_s\sum_i \frac{|T_{p_i}|}{V_i}\ll \!{(\gtrsim)}\, 1$ (note the absolute values in the sum) and as positive (negative) if $g_s\sum_i \frac{T_{p_i}}{V_i} > \!{(<)\,} 0$.

\paragraph{No sources or small source terms}~\\
Let us first consider the case where we have either no sources ($T_{p_i}=0$) or only small (positive or negative) source terms. 
The source terms do then not contribute at leading order in $D$ to \eqref{f1}, \eqref{f2} since $g_s\sum_i \frac{|T_{p_i}|}{V_i}\ll 1$ implies $g_s|\sum_i \frac{T_{p_i}}{V_i}| \ll 1$ and $g_s|\sum_i p_i \frac{T_{p_i}}{V_i}| \ll D$.
It thus follows from \eqref{f1} that $\mathcal{R}_{D-d} \gtrsim D$. Furthermore,  \eqref{f2} requires either $|H_3|^2 \gtrsim D$ or $g_s^2\sum_q|F_q|^2 \gtrsim 1$ with sufficiently large $\overline{q}>D/2$. In the second case, \eqref{mn} also implies $|\mathcal{R}_d|\gtrsim d$.

Depending on the sign of the vacuum energy, there are further constraints. In particular, \eqref{mn} implies that having $g_s^2\sum_q|F_q|^2 \gtrsim 1$ or positive source terms is not compatible with dS or Minkowski. In the case of no sources, Minkowski (for $F_q=0$) but no dS is possible, again due to \eqref{mn}.
For negative source terms, Minkowski and dS are a priori not ruled out by \eqref{mn}, \eqref{f1}, \eqref{f2}.
We remark, however, that a more detailed analysis in Section \ref{sec:b3} reveals a conflict between small source terms and tadpole cancellation for dS and Minkowski vacua in a large class of models. For AdS vacua, analogous issues can be shown to arise in the case where $g_s^2\sum_q|F_q|^2 \ll 1$.

\paragraph{Large positive source terms}~\\
We now consider the case where the source terms are large ($g_s\sum_i \frac{|T_{p_i}|}{V_i} \gtrsim 1$) and positive ($g_s\sum_i \frac{T_{p_i}}{V_i} > 0$).
\eqref{f1} then again implies $\mathcal{R}_{D-d} \gtrsim D$. Furthermore, if $\overline{p}\ge (D-4)/2$, \eqref{f2} requires either $|H_3|^2 \gtrsim D$ or $g_s^2\sum_q|F_q|^2 \gtrsim 1$ with sufficiently large $\overline{q}>D/2$. In the last case, we further conclude that $|\mathcal{R}_d|\gtrsim d$ due to \eqref{mn}. It also follows from \eqref{mn} that neither dS nor Minkowski vacua are possible when the source terms are positive.

\paragraph{Large negative source terms}~\\
We finally consider the case where the source terms are large ($g_s\sum_i \frac{|T_{p_i}|}{V_i} \gtrsim 1$) and negative ($g_s\sum_i \frac{T_{p_i}}{V_i} < 0$). Recall that \eqref{stab} then imposes that, for dS/Minkowski, the absolute value of the source terms is not parametrically larger than the RR energy densities:
\begin{equation}
g_s \Big|\sum_i \frac{T_{p_i}}{V_i}\Big| \lesssim g_s^2\sum_q |F_q|^2. \label{dilstab}
\end{equation}
For AdS, the same conclusion follows from \eqref{mn}.

There are now two cases to consider: $g_s |\sum_i \frac{T_{p_i}}{V_i}| \ll D/d$ and $g_s |\sum_i \frac{T_{p_i}}{V_i}| \gtrsim D/d$.
Note that $g_s |\sum_i \frac{T_{p_i}}{V_i}|$ can in principle be very small in spite of $g_s\sum_i \frac{|T_{p_i}|}{V_i} \gtrsim 1$ if there are parametric cancellations between positive-tension and negative-tension sources. In the first case, we can neglect the source terms in \eqref{f1} at leading order in $D$. Satisfying \eqref{f1} then requires $\mathcal{R}_{D-d}\gtrsim D$. In the second case, no constraint on $\mathcal{R}_{D-d}$ follows from \eqref{f1}. However, \eqref{dilstab} then imposes $g_s^2\sum_q |F_q|^2 \gtrsim D/d$. Furthermore, in both cases, \eqref{f2} for $\overline{p}\le(D-4)/2$ requires either $|H_3|^2 \gtrsim D$ or $g_s^2\sum_q |F_q|^2\gtrsim 1$ with sufficiently large $\overline{q}>D/2$.

\bigskip

We summarized in Table \ref{tab1} the various lower bounds on the curvature/energy densities implied by the above arguments.
We stress again that we only imposed the dilaton equation and the external and traced internal Einstein equations (or, equivalently, the equation for the volume modulus) as well as the requirement that the dilaton is not a tachyon. Our bounds are therefore necessary but not sufficient conditions for the existence of a solution. Further constraints may arise, e.g., from the stabilization of the non-universal moduli and from tadpole cancellation conditions.
We will not present an exhaustive analysis of such model-dependent constraints here but only give a simple example.

Consider a collection of parallel O$p$-planes whose tadpole is cancelled by a flux term $H_3 \w F_{D-p-4}$. Let us also assume for simplicity that $H_3$ and $F_{D-p-4}$ do not have any pieces with vanishing wedge product. As will be derived in Section \ref{sec:b3}, this implies the constraint
\begin{equation}
g_s^2 |F_{D-p-4}|^2 |H_3|^2 = \left(g_s \sum_{i} \frac{Q_{p_i}}{2\pi V}\right)^2,
\end{equation}
where $V$ is the volume of the transverse space of the O-planes and $Q_{p_i}=T_{p_i}$ are their RR charges.
We thus see that tadpole cancellation relates the energy densities $g_s^2 |F_{D-p-4}|^2$ and $|H_3|^2$ to the source terms $g_s \sum_{i} \frac{|T_{p_i}|}{V}$,
which can lead to further constraints beyond those listed in Table \ref{tab1}.
In Section \ref{sec:b3}, we will study such constraints more carefully than here and argue in particular that supercritical dS vacua cannot have small source terms ($g_s \sum_{i} \frac{|T_{p_i}|}{V_i}\ll 1$).

\bigskip

\begin{table}[t]\renewcommand{\arraystretch}{1.2}\setlength{\tabcolsep}{6pt}
\hspace{-1.2cm}
\begin{tabular}{|l|l|l|l|l|l|}
\hline 
sources & $\overline{p}$ & $\overline{q}$ & AdS & Minkowski & dS \\
\hline 
\multirow{3}{*}{no sources} & & $\le \frac{D}{2}$ & $\mathcal{R}_{D-d}\gtrsim D$, $|H_3|^2\gtrsim D$ & \multirow{3}{*}{$\mathcal{R}_{D-d}\gtrsim D$, $|H_3|^2\gtrsim D$ } & \multirow{3}{*}{ruled out} \\
\cline{3-4}
 & & \multirow{2}{*}{$>\frac{D}{2}$} & $\mathcal{R}_{D-d}\gtrsim D$, $|H_3|^2\gtrsim D$ or &  & \\
           & & & $\mathcal{R}_{D-d}\gtrsim D$, $g_s^2\sum_q|F_q|^2\gtrsim 1$, $|\mathcal{R}_d|\gtrsim d$ & &\\
\hline
\multirow{3}{*}{small pos.} & & $\le \frac{D}{2}$ & $\mathcal{R}_{D-d}\gtrsim D$, $|H_3|^2\gtrsim D$ & \multicolumn{2}{ |l| }{\multirow{3}{*}{ruled out}} \\
\cline{3-4}
 & & \multirow{2}{*}{$>\frac{D}{2}$} & $\mathcal{R}_{D-d}\gtrsim D$, $|H_3|^2\gtrsim D$ or &  \multicolumn{2}{ |l| }{} \\
           & & & $\mathcal{R}_{D-d}\gtrsim D$, $g_s^2\sum_q|F_q|^2\gtrsim 1$, $|\mathcal{R}_d|\gtrsim d$ & \multicolumn{2}{ |l| }{} \\
\hline
\multirow{3}{*}{small neg.} & & $\le \frac{D}{2}$ & $\mathcal{R}_{D-d}\gtrsim D$, $|H_3|^2\gtrsim D$ &  \multicolumn{2}{ |l| }{\multirow{3}{*}{$\mathcal{R}_{D-d}\gtrsim D$, $|H_3|^2\gtrsim D$}} \\
\cline{3-4}
 & & \multirow{2}{*}{$>\frac{D}{2}$} & $\mathcal{R}_{D-d}\gtrsim D$, $|H_3|^2\gtrsim D$ or &   \multicolumn{2}{ |l| }{} \\
           & & & $\mathcal{R}_{D-d}\gtrsim D$, $g_s^2\sum_q|F_q|^2\gtrsim 1$, $|\mathcal{R}_d|\gtrsim d$ &  \multicolumn{2}{ |l| }{} \\
\hline
\multirow{4}{*}{large pos.} & $< \frac{D-4}{2} $ & & $\mathcal{R}_{D-d}\gtrsim D$   &  \multicolumn{2}{ |l| }{\multirow{4}{*}{ruled out}} \\
\cline{2-4}
&\multirow{3}{*}{$\ge \frac{D-4}{2} $} & $\le \frac{D}{2}$ & $\mathcal{R}_{D-d}\gtrsim D$, $|H_3|^2\gtrsim D$ &  \multicolumn{2}{ |l| }{} \\
\cline{3-4}
&  & \multirow{2}{*}{$>\frac{D}{2}$} & $\mathcal{R}_{D-d}\gtrsim D$, $|H_3|^2\gtrsim D$ or &  \multicolumn{2}{ |l| }{} \\
           & & & $\mathcal{R}_{D-d}\gtrsim D$, $g_s^2\sum_q|F_q|^2\gtrsim 1$, $|\mathcal{R}_d|\gtrsim d$ &  \multicolumn{2}{ |l| }{} \\
\hline
\multirow{3}{*}{large neg.}  & \multirow{2}{*}{$\le \frac{D-4}{2} $} & $\le \frac{D}{2}$ & \multicolumn{3}{ |l| }{$\mathcal{R}_{D-d}\gtrsim D$, $|H_3|^2\gtrsim D$ or $g_s^2\sum_q|F_q|^2\gtrsim \frac{D}{d}$, $|H_3|^2\gtrsim D$} \\
\cline{3-6}
 &  & $>\frac{D}{2}$ & \multicolumn{3}{ |l| }{$\mathcal{R}_{D-d}\gtrsim D$, $|H_3|^2\gtrsim D$ or $\mathcal{R}_{D-d}\gtrsim D$, $g_s^2\sum_q|F_q|^2\gtrsim 1$ or $g_s^2\sum_q|F_q|^2\gtrsim \frac{D}{d}$} \\
\cline{2-6}
& $> \frac{D-4}{2}$ & & \multicolumn{3}{ |l| }{$\mathcal{R}_{D-d}\gtrsim D$ or $g_s^2\sum_q|F_q|^2\gtrsim \frac{D}{d}$} \\
\hline
\end{tabular} 
\caption{Lower bounds on string-frame curvature/energy densities imposed by \eqref{mn}, \eqref{stab}, \eqref{f1}, \eqref{f2}.
These are necessary but not sufficient conditions for the existence of a solution (see, e.g., Section \ref{sec:b3} for additional constraints from tadpole cancellation). Source terms are denoted as small (large) if $g_s\sum_i \frac{|T_{p_i}|}{V_i}\ll \!{(\gtrsim)}\, 1$ and as positive (negative) if $g_s\sum_i \frac{T_{p_i}}{V_i} > \!{(<)\,} 0$.
}

\label{tab1}
\end{table}

The discussion in this section so far applied to geometric compactifications, i.e., assuming that the metric of the compact space is dynamical and satisfies the Einstein equations.
Let us now also discuss a different class of models where this is not the case. In particular, one of the earliest proposals for a dS model in string theory is an asymmetric toroidal orbifold where all NSNS moduli except for the dilaton are projected out \cite{Silverstein:2001xn, Maloney:2002rr}. This implies that $\mathcal{R}_{D-d}=|H_3|^2=0$ and that there is no equation of motion or stability condition for the volume modulus. The only relevant constraints are thus the dilaton equation \eqref{f1} and the dilaton stability condition \eqref{stab}, which with \eqref{mn} implies \eqref{dilstab} for dS.
Since $\mathcal{R}_{D-d}=|H_3|^2=0$, \eqref{f1} yields $-g_s\sum_i \frac{T_{p_i}}{V_i}\gtrsim D/d$. Using this together with \eqref{dilstab}, we conclude
\begin{equation}
g_s^2\sum_q |F_q|^2\gtrsim \frac{D}{d}.
\end{equation}
We thus find that the RR energy density is large in such models, analogously to our results for the geometric models.

\subsection{Tension with suppression of $\alpha^\prime$ corrections?}
\label{sec:string2}

We have seen that supercritical compactifications necessarily have large string-frame curvature and/or energy densities. As we now explain, this may be in tension with the requirement of suppressing $\alpha^\prime$ corrections to the classical effective action \eqref{action}.

To see this, recall that the effective action receives an infinite number of higher-derivative corrections at string tree and loop levels. This includes in particular terms involving powers of the curvature and RR/NSNS fluxes, which are of the schematic form
\begin{equation}
\delta\mathcal{L} \supset g_s^{2l-2} \left(\mathcal{R}_{\circ\circ}{}^\circ{}_\circ\right)^m\, \left(H_{3\circ\circ\circ}\right)^{2n}\,
\left(\prod\limits_q\left(g_s F_{q\circ\circ\cdots\circ}\right)^{2r_q}\right) \left(g^{\circ\circ}\right)^{m+3n+\sum_q qr_q},
\label{hdt}
\end{equation}
with $m,n,l,r_q\in\mathbb{N}_0$.\footnote{More generally, there can also be terms where covariant derivatives act on some of the above factors and terms involving derivatives of the dilaton. Such corrections will not be relevant for our argument. Also note that, in the critical type II string theories, the earliest (bulk) higher-derivative terms are believed to arise at the 8-derivative order, i.e., $m+n+\sum_q r_q\ge 4$.} The circle symbols stand for the $D$-dimensional spacetime indices of the various tensors, which can be contracted in many different ways. Note that the $\alpha^\prime$ expansion for the RR fields is not an expansion in powers of $F_q$ but in powers of $g_sF_q$. This is related to an extra factor of $g_s$ in the corresponding vertex operators \cite{Polchinski:1998rr, DeWolfe:2005uu}.\footnote{One might therefore call the corresponding higher-derivative terms ``loop corrections'' rather than ``$\alpha^\prime$ corrections''. In this paper, we do not care about this distinction and will refer to all higher-derivative terms as $\alpha^\prime$ or string corrections.}

The classical approximation is valid as long as the higher-derivative terms are suppressed compared to the two-derivative ones.\footnote{There are special backgrounds like the linear-dilaton theory which are exact in $\alpha^\prime$ \cite{Myers:1987fv, Antoniadis:1988aa, deAlwis:1988pr, Polchinski:1998rq, Polchinski:1998rr}. Such an all-order cancellation is not expected for general flux vacua
and would again have to be established by arguments that go beyond studying the classical effective action.
}
Assuming $\mathcal{O}(1)$ expansion coefficients,
\eqref{hdt} thus suggests that one should demand
\begin{equation}
|\mathcal{R}_d|, |\mathcal{R}_{D-d}|, |H_3|^2, g_s^2|F_q|^2 \ll 1
\end{equation}
as a necessary condition for control over the $\alpha^\prime$ expansion. Indeed, this is typically imposed in the critical string theories.
If this condition is also correct in the supercritical case, it is in obvious tension with the results of Section \ref{sec:string1}, which we summarized in Table \ref{tab1}.

On the other hand, as suggested in \cite{Harribey:2018xvs, DeLuca:2021pej, Flauger:2022hie}, the above conclusion may be too premature at large $D$. In particular, one may speculate that the expansion coefficients of the higher-derivative terms are functions of $D$ which ensure a large-$D$ suppression in spite of the large curvature and energy densities. One should further keep in mind that two terms constructed from the same tensors
can scale very differently with $D$ depending on their precise index structure. This is the case, for example, for the various $\mathcal{R}^4$ curvature invariants, as we will see below. Both effects may contribute to suppressing the higher-derivative terms more than naively expected. If this is true, the large curvature and energy densities found in Section \ref{sec:string1}
might be ok at sufficiently large $D$.

As a simple example, consider a compactification with large RR fluxes, $g_s^2|F_q|^2\gtrsim D$, as required in many models according to the results of Section \ref{sec:string1}.
For concreteness, we furthermore consider the higher-derivative term
\begin{equation}
\delta \mathcal{L} = c_1 g_s^{-2} \left( g_s^2|F_q|^2 \right)^4. \label{f8}
\end{equation}
For $c_1\sim\mathcal{O}(1)$, this is parametrically larger than the two-derivative term $|F_q|^2$, indicating that the classical effective action \eqref{action} is no longer trustworthy. On the other hand, if $|c_1| \ll \frac{1}{D^3}$, \eqref{f8} would be negligible in spite of the large RR field strength.
Indeed, it was proposed in \cite{DeLuca:2021pej, Flauger:2022hie} (citing unpublished work) that a parametric suppression of higher-derivative RR terms does occur at large $D$.

As another example, consider a compactification on a $(D-d)$-sphere with radius $R$ and $d\ll D$. The curvature tensors are
\begin{equation}
\mathcal{R}_{mnpq} = \frac{1}{R^2}(g_{mp}g_{nq}-g_{mq}g_{np}), \quad \mathcal{R}_{mn} = \frac{D-d-1}{R^2} g_{mn}, \quad
\mathcal{R}_{D-d} = \frac{(D-d)(D-d-1)}{R^2}.
\end{equation}
Let us assume $R^2\sim D$ such that $\mathcal{R}_{D-d}\sim D$, as required in many models according to Section \ref{sec:string1}.
Now consider the following curvature corrections at the 8-derivative order:
\begin{equation}
\delta \mathcal{L} = g_s^{-2}\left(c_1 \mathcal{R}_{D-d}^4
+ c_2 (\mathcal{R}_{mn}\mathcal{R}^{mn})^2
+ c_3 (\mathcal{R}_{mnpq}\mathcal{R}^{mnpq})^2 + \ldots\right),
\end{equation}
where the dots stand for $\mathcal{R}^4$ terms involving other ways to contract the Riemann tensors. Using the above formulae, we find
\begin{align}
\mathcal{R}_{D-d}^4&\sim D^4, \qquad
(\mathcal{R}_{mn}\mathcal{R}^{mn})^2 \sim D^2, \qquad
(\mathcal{R}_{mnpq}\mathcal{R}^{mnpq})^2 \sim D^0.
\end{align}
We thus see that the large-$D$ scaling of the $\mathcal{R}^4$ terms crucially depends on their index structure. It follows that the $\mathcal{R}^4$ corrections are under control in spite of large $\mathcal{R}_{D-d}\sim D$ if the coefficients $c_i$ satisfy $|c_1|\ll \frac{1}{D^3}$, $|c_2|\ll \frac{1}{D}$, $|c_3|\ll D$.

\bigskip

To summarize, we showed that the equations of motion and stability requirements impose lower bounds on NSNS/RR energy densities and curvature invariants, stated in Table \ref{tab1}. This implies that higher-derivative corrections to the classical effective action are only under control if their coefficients do not exceed (model-dependent) upper bounds which scale non-trivially with $D$. It would be very interesting to check whether these bounds are satisfied in string theory, e.g., by computing string amplitudes at large $D$.

Importantly, even if the smeared solutions considered in this section do not have large string corrections, one should still check that backreaction effects do not change this conclusion. In particular, O-planes have the effect of generating singular holes in their vicinity. If these holes are too large, the classical description breaks down even when the smeared solution naively suggests no problem with string corrections. Such backreaction effects will be discussed in Section \ref{sec:backreaction}.

\subsection{Curvature and the KK scale}
\label{sec:string3}

Another important consistency condition for the validity of the $D$-dimensional effective action is that the KK scale is below the scale of the massive string states. 
We will discuss this in more detail in Section \ref{sec:b5} and be brief here.
The KK scale associated to a compact space is determined by the first eigenvalue $\lambda_1$ of the Laplacian.
Given an $n$-dimensional manifold with bounded Ricci curvature such that $\mathcal{R}_{mn}u^mu^n \ge (n-1)K$ for some $K>0$ and all unit tangent vectors $u^m$, we can use the Lichnerowicz theorem, which states that $\lambda_1 \ge nK$ \cite{Lichnerowicz1958}. Consider, for example, an Einstein space with positive $\mathcal{R}_{D-d}\sim D$. We can then write
\begin{equation}
\lambda_1 \ge \frac{\mathcal{R}_{D-d}}{D-d-1}.
\end{equation}
We thus find $\lambda_1 \gtrsim 1$ if the compact space is high-dimensional ($d\ll D$). Massive string states are thus comparable to or lighter than the KK scale unless they satisfy $M_s^2 \gg 1$. An even stronger bound $\lambda_1 \gtrsim D$ is obtained for low-dimensional compact spaces ($D-d=\mathcal{O}(1)$). The masses of the string states thus drop below the KK scale in such backgrounds unless they are parametrically large, $M_s^2\gg D$.

As another example, consider a compact space which is a product of $\mathcal{O}(D)$ low-dimensional Einstein spaces. Assuming again $\mathcal{R}_{D-d}\sim D$, at least one of these spaces must have a positive scalar curvature $\gtrsim\mathcal{O}(1)$.
The Lichnerowicz theorem thus again yields $\lambda_1\gtrsim 1$ for KK modes on this Einstein space. It would be interesting to study more systematically how the Einstein equations constrain the individual components of $\mathcal{R}_{mn}$ (and thus $K$) for more general spaces.
We will not attempt to do this and instead continue our discussion of bounds on the KK scale in Section \ref{sec:b5}. We will in particular see there that the O-plane backreaction in dS models puts further constraints on the KK scale beyond those discussed here.
Interestingly, these constraints do not require a positively curved space but also arise in spaces with zero or negative curvature.

\section{Backreaction}
\label{sec:backreaction}

In this section, we study the backreaction of localized sources in string theories with general $D$. One of our key observations is that the backreaction of the O-planes required in supercritical dS models creates singular holes which eat up a large fraction of the classical spacetime.

In Section \ref{sec:b1}, we compute the leading backreaction corrections for general sources (D-branes and O-planes) and explain that, for O-planes, a large backreaction is tied to a breakdown of the classical solution.
We find that the backreaction is small if
\begin{equation}
D g_s \sum_i |T_{p_i}| G_i \ll 1 \label{i1}
\end{equation}
up to $\mathcal{O}(1)$ factors, where $G_i$ is the Green's function of the Laplacian of the space transverse to the corresponding source.
In Section \ref{sec:b2}, we estimate the Green's functions as
\begin{equation}
G_i \gtrsim \frac{\mathcal{D}_i^2}{n_i V_i},
\end{equation}
where $n_i=D-p_i-1$ is the dimension, $\mathcal{D}_i$ is the string-frame diameter and $V_i$ is the volume of the transverse space.
The discussion in Sections \ref{sec:b1} and \ref{sec:b2} applies to general compactifications in critical and non-critical string theories.

In Section \ref{sec:b3}, we then specialize to supercritical dS models and argue that the source terms must be large in such models,
\begin{equation}
g_s \sum_i \frac{|T_{p_i}|}{V_i} \gtrsim 1. \label{i3}
\end{equation}
Readers not interested in the arguments leading to the inequalities \eqref{i1}--\eqref{i3} may skip Sections \ref{sec:b1}--\ref{sec:b3} and proceed directly to Section \ref{sec:b4} where we use them to derive one of our main results. In particular, we argue there that controlling the O-plane backreaction in supercritical dS vacua would require the average diameter of the transverse spaces to be small in string units,
\begin{equation}
\overline{\frac{\mathcal{D}^2}{n}} \ll \frac{1}{D},
\end{equation}
and we show that this is often in conflict with rigorous geometric bounds or with flux quantization. In Section \ref{sec:b5}, we furthermore argue that the required small diameters are in tension with the requirement that massive string states should not be lighter than the KK scale.

\begin{figure}[t!]
\centering
\includegraphics[trim = 0mm 10mm 0mm 20mm, clip, width=\textwidth]{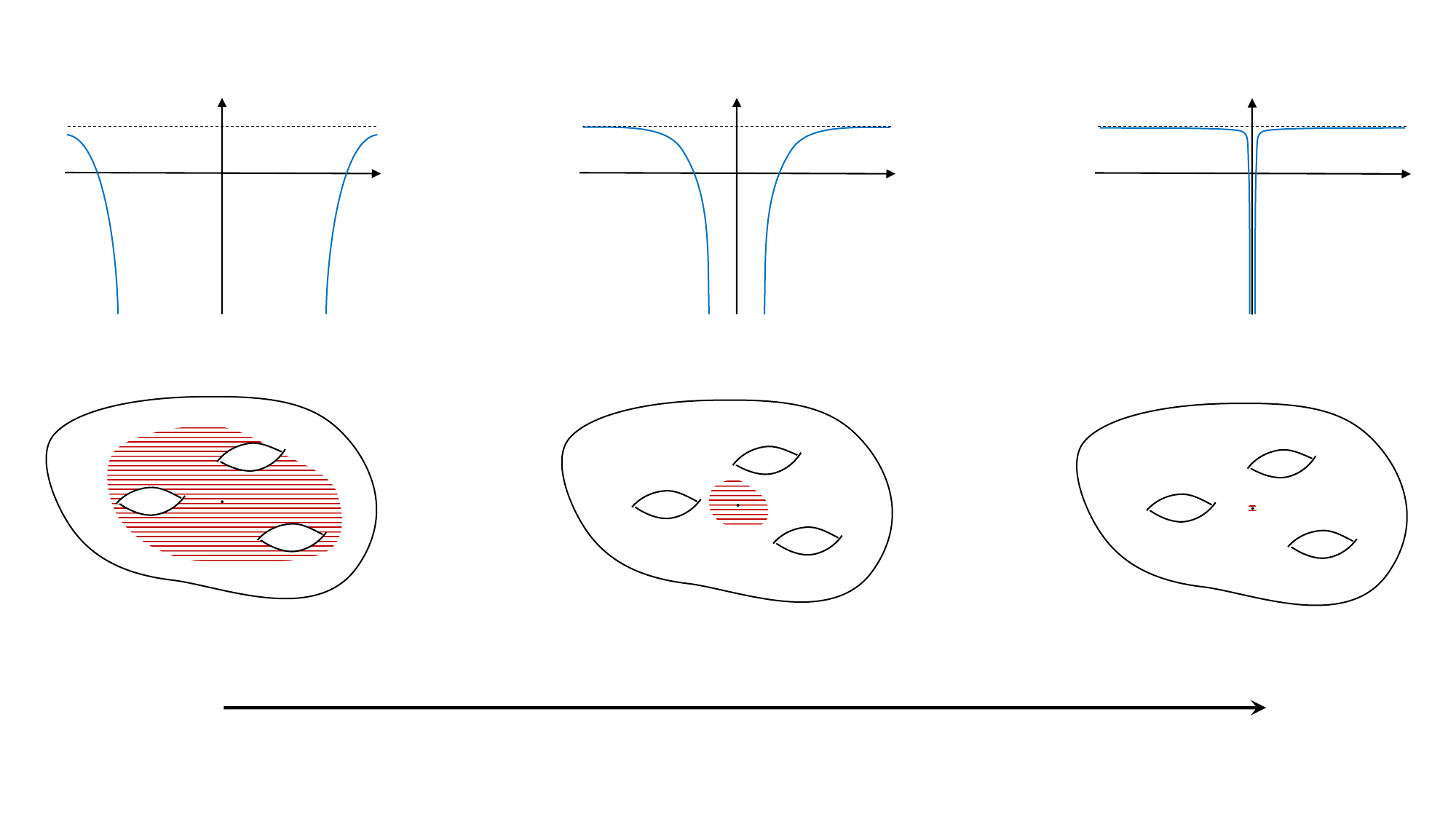}
\put(-230,6){$\scriptstyle{Dg_s|T_p| G \to\, 0}$}

\caption{Schematic depiction of the smeared limit on the space transverse to an O-plane.
In the region where the O-plane backreacts strongly (red shade), fields like the warp factor or the dilaton (blue lines) become singular, signaling that the classical description in terms of a $D$-dimensional effective field theory cannot be trusted. As we take the limit $Dg_s|T_p|G\to 0$, the backreaction is confined to a smaller and smaller fraction of the transverse space until the field profiles agree with their smeared values (dashed lines) everywhere. In this regime, the classical solution
is reliable.
\label{smeared}}
\end{figure}

\subsection{Backreaction and singularities}
\label{sec:b1}

Many flux compactifications in the literature rely on the so-called smeared approximation. Technically, this means that the localized source terms $\sim \delta(\Sigma_i)$ in the equations of motion are replaced by constants, i.e., the charge and energy densities of the branes or O-planes are assumed to be effectively spread out over the whole transverse space.

Since there are often misconceptions about smearing in the literature, let us emphasize that smearing does \emph{not} require the sources to be physically delocalized. Rather, the smeared solution should be understood as a computational device describing an effective behavior at long distances. Indeed, the smeared solution arises as the leading-order term in a systematic expansion of the fully backreacted solution in a regime where only the zero mode of the source backreacts at long distances  \cite{Junghans:2020acz, Marchesano:2020qvg}. The backreaction of all other modes is small in this regime and can be treated as a linear perturbation.
The source thus effectively appears smeared at long distances but its localized nature can still be probed at sufficiently small distances.
In particular, the often made claim that O-planes cannot be smeared because they are intrinsically localized is false.

Quite the opposite, we claim that the existence of a smeared regime is a \emph{crucial} requirement in order that a classical solution with O-planes can make sense at all! The reason is that O-planes are known to generate singular ``holes'', i.e., finite-distance singularities
where the description in terms of the classical equations of motion breaks down and string corrections blow up. As we will explain below, these unreliable regions shrink to zero in the smeared limit and the classical description becomes reliable.
On the other hand, in a solution which is not in the smeared regime, a large part of the compactification space is eaten up by the singular holes and there is no reason to believe that the classical solution is meaningful, see Fig.~\ref{smeared}. For consistency, we should therefore avoid the regime where the O-planes backreact non-linearly on a large part of the spacetime and consider instead a regime where the backreaction is small except at very small distances.

Two well-known examples of flux compactifications with O-planes where the above claims were explicitly verified are DGKT AdS vacua \cite{DeWolfe:2005uu, Camara:2005dc} in type IIA string theory and GKP Minkowski vacua \cite{Giddings:2001yu} in type IIB string theory. For DGKT, it was explicitly shown in \cite{Junghans:2020acz, Marchesano:2020qvg} that the fully backreacted solution approaches the smeared one at large $4$-form flux (see also \cite{Cribiori:2021djm, Emelin:2022cac} for similar results in related setups). For GKP, an analogous behavior at large volume was shown in \cite[App.~A]{Junghans:2020acz}. See also \cite{Blaback:2010sj} for earlier work and \cite{Baines:2020dmu} for a pedagogical discussion of these ideas.

In the following, we provide a more general discussion of backreaction corrections which in particular applies to general D$p$-branes/O$p$-planes and general dimension $D$. Our discussion follows the general approach developed in \cite{Junghans:2020acz} and we refer the reader to that paper for more details.

\subsubsection{Leading corrections}
\label{sec:single}

In general, finding solutions to the classical equations of motion including the full non-linear backreaction of the various sources is extremely difficult, as it involves solving a system of coupled non-linear 2nd-order PDEs (1st order for supersymmetric solutions). However, in a region of the spacetime where the backreaction is small,
we can expand
\begin{equation}
\varphi = \epsilon^a \sum_{b=0}^\infty \epsilon^b \varphi_b, \label{exp}
\end{equation}
where $\varphi\in\{\e^A, \e^\phi, g_{mn},C_{q-1},B_2\}$ denotes any field,
$\tilde\varphi\equiv\epsilon^a\varphi_0$ is the smeared solution and
$\varphi_1,\varphi_2,\ldots$ are higher-order corrections encoding backreaction effects. The exponent $a$ can in general be different for each field (or field component).
As we will see below, the leading corrections $\varphi_1$ are determined by simple Poisson equations.
The expansion parameter $\epsilon$ is for now just a formal device to organize the expansion and we will have to justify a posteriori that this expansion is self-consistent. We will discuss the corresponding condition below. For example, in DGKT, one finds $\epsilon=1/N$, where $N$ is related to the 4-form fluxes, so that expanding around the smeared solution is justified in the large-flux regime \cite{Junghans:2020acz, Marchesano:2020qvg}. Since $g_s \sim N^{-3/4}$, $\mathcal{V} \sim N^{3/2}$ (where $\mathcal{V}$ is the Calabi-Yau volume in the string frame), this is at the same time a regime of small string coupling and large volume.

Our goal is now to compute the corrections $\varphi_1$ explicitly for general $D$.
For simplicity, we consider a single O$p$-plane of codimension $n=D-p-1$ and tension $\frac{T_p}{2\pi}=-1$. The case of several sources is straightforwardly obtained by superposition and will be discussed further below.

Instead of computing perturbative solutions of the form \eqref{exp} in specific models, we try to remain as general as possible. In particular, there can be many different scaling limits of the various fields in which the backreaction becomes small, and we do not wish to analyze all of them case by case (see \cite{Junghans:2020acz, Marchesano:2020qvg, Cribiori:2021djm, Emelin:2022cac} for explicit examples). Our only input in the following is to assume that we are in a regime where a) corrections to (components of) fields which are non-zero in the smeared solution are small and b)
terms in the equations of motion which are quadratic or higher in the corrections can be neglected.\footnote{Note that the second assumption does not follow from the first one since many compactifications have fields which are zero in the smeared solution but sourced by the backreaction.}
Under these two assumptions, the leading corrections satisfy a set of linear equations which can be solved explicitly and are derived from  \eqref{eom1}--\eqref{eom3}, \eqref{eom4} as follows:
\begin{itemize}
\item We first note that the dilaton and the warp factor are constant in the smeared solution (with $\tilde\phi = \ln g_s$ and $\tilde A=0$). This implies that the terms $(\partial\phi)^2$, $(\partial A)^2$ and $(\partial A \cdot \partial\phi)$ in \eqref{eom1}--\eqref{eom3} are at least quadratic in the backreaction corrections and can therefore be discarded. The second-derivative terms $\e^{\phi}\nabla^2\e^{-\phi}$ and $\e^{-A}\nabla^2\e^{A}$ contain terms which are linear in the corrections and are therefore kept. They reduce to $g_s\tilde\nabla^2\e^{-\phi}$ and $\tilde\nabla^2\e^{A}$ at leading order, where $\tilde\nabla^2$ is the Laplacian with respect to the smeared metric $\tilde g_{mn}$.

\item Expanding the Ricci tensor around the smeared solution yields \cite{Junghans:2020acz}
\begin{equation}
\mathcal{R}_{mn}= \mathcal{\tilde R}_{mn} + \frac{1}{2} \tilde g^{rs}\left( \tilde\nabla_s \tilde\nabla_m g_{rn} + \tilde\nabla_s \tilde\nabla_n g_{rm} \right) - \frac{1}{2} \tilde g^{rs}\tilde\nabla_m \tilde\nabla_n g_{rs} - \frac{1}{2} \tilde\nabla^2 g_{mn}
\end{equation}
up to terms quadratic or higher in the backreaction corrections, where $\mathcal{\tilde R}_{mn}$ and $\tilde\nabla_m$ are the Ricci tensor and the covariant derivative with respect to $\tilde g_{mn}$. Here we used that terms of the form $(\tilde\nabla g)^2$ are at least quadratic in the backreaction corrections since the smeared metric satisfies $\tilde\nabla\tilde g=0$ by definition of $\tilde\nabla$.

\item The remaining terms in \eqref{eom1}--\eqref{eom3} and \eqref{eom4} are simplified by discarding the subleading corrections to the smeared fields. For example, we can replace $\e^{2\phi}|F_q|^2$ and $|H_3|^2$
by $g_s^2|\tilde F_q|^2$ and $|\tilde H_3|^2$
at leading order and analogously for other terms, where the tildes indicate that the objects are built using the smeared solution. We also replace $H_3\w F_{n-3}$ in \eqref{eom4} by $\tilde H_3 \w \tilde F_{n-3}$ at leading order.\footnote{If $H_3$ or $F_{n-3}$ have several components, the two assumptions stated above do a priori not exclude that some of the backreaction corrections in $H_3\w F_{n-3}$ contribute at the same order as the smeared term $\tilde H_3\w \tilde F_{n-3}$. However, we are not aware of examples of this kind and will assume a regime where this does not happen.
}
\end{itemize}
After these steps, the equations of motion reduce to the following set of equations:
\begin{align}
0 &= -8 g_s \tilde\nabla^2 \e^{-\phi} -4d \tilde\nabla^2 \e^A + 2 \mathcal{ R}_d + 2 \tilde{\mathcal{R}}_{D-d} - 2\tilde g^{mn}\tilde \nabla^2 g_{mn} + 2\tilde\nabla^m \tilde\nabla^n g_{mn} - 8\pi^2 (D-10) \nl - |\tilde H_3|^2  + g_s \tilde\delta(\Sigma),  \label{e10} \\
0 &=  \frac{2d}{D-2}g_s\tilde\nabla^2\e^{-\phi} +d\tilde\nabla^2\e^A - \mathcal{R}_d + \frac{d}{D-2} 4\pi^2 (D-10) - \frac{d}{D-2} |\tilde H_3|^2 \nl - \sum_{q} \frac{(q-1)d}{2(D-2)}g_s^2 |\tilde F_q|^2 + \frac{(n-2)d}{2(D-2)} g_s\tilde \delta(\Sigma), \\
0 &= 2g_s \tilde \nabla_m\partial_n \e^{-\phi} + d\tilde\nabla_m\partial_n \e^A + \frac{2}{D-2} \tilde g_{mn}g_s\tilde \nabla^2\e^{-\phi} -\tilde{\mathcal{R}}_{mn}- \frac{1}{2} \tilde g^{rs}\left( \tilde\nabla_s \tilde\nabla_m g_{rn} + \tilde\nabla_s \tilde\nabla_n g_{rm} \right) \nl + \frac{1}{2} \tilde g^{rs}\tilde\nabla_m \tilde\nabla_n g_{rs} +  \frac{1}{2} \tilde\nabla^2 g_{mn}
 + \tilde g_{mn}\frac{4\pi^2}{D-2} (D-10) + \frac{1}{2}|\tilde H_3|^2_{mn} - \frac{1}{D-2}\tilde g_{mn}|\tilde H_3|^2 \nl + \frac{1}{2} g_s^2 \sum_{q} \left( |\tilde F_q|^2_{mn}- \frac{q-1}{D-2}\tilde g_{mn}|\tilde F_q|^2 \right) + \frac{1}{2} \left(\tilde \Pi_{mn}-\frac{D-n}{D-2}\tilde g_{mn}\right) g_s \tilde\delta(\Sigma),  \label{e30} \\
0 &= -\d F_{n-1} + \tilde H_3 \w \tilde F_{n-3} \pm (-1)^{n(n-1)/2} \tilde\delta(\Sigma) \tilde{\text{dvol}}_{n}, \label{e40}
\end{align}
where $\tilde{\text{dvol}}_{n}=\tilde\star_n 1$ is the volume form of the transverse space and
tildes indicate as before that we use the smeared fields. The upper sign in \eqref{e40} applies to O-planes and the lower one to anti-O-planes.

The final step is to substitute the smeared fields in each equation with the smeared source term $\sim 1/\tilde V$ using the smeared equations of motion (i.e., those where the delta distributions are replaced by \eqref{smear}), where $\tilde V$ is the volume of the transverse space.
We thus arrive at\footnote{One might be worried that there could be parametric cancellations between the various smeared terms in \eqref{e10}--\eqref{e30} such that $g_s/\tilde V$ is very small and competes with the backreaction corrections we discarded earlier. A general analysis of this possibility is beyond the scope of this work. However, we will find in Section \ref{sec:b3} that, in the dS models studied in this paper, the source term cannot be arbitrarily small but satisfies $g_s/\tilde V \gtrsim 1$. Some checks we performed in examples suggest that the discarded backreaction corrections are then subleading compared to $g_s/\tilde V$ and can self-consistently be neglected in the regime $Dg_sG\ll 1$ (where $G$ will be defined below), which is consistent with independent arguments in Section \ref{sec:val} leading to the same condition.
}
\begin{align}
0 &= -8 g_s \tilde\nabla^2 \e^{-\phi} -4d \tilde\nabla^2 \e^A - 2\tilde g^{mn}\tilde \nabla^2 g_{mn} + 2\tilde\nabla^m \tilde\nabla^n g_{mn} - g_s \left(\frac{1}{\tilde V}-\tilde\delta(\Sigma)\right), \label{e1} \\
0 &=  \frac{2d}{D-2}g_s \tilde\nabla^2\e^{-\phi} +d \tilde\nabla^2\e^A  -  \frac{(n-2)d}{2(D-2)} g_s \left(\frac{1}{\tilde V}-\tilde\delta(\Sigma)\right), \\
0 &= 2g_s \tilde\nabla_m\partial_n \e^{-\phi} + d \tilde\nabla_m\partial_n \e^A + \frac{2}{D-2} \tilde g_{mn} g_s \tilde\nabla^2\e^{-\phi} - \frac{1}{2} \tilde g^{rs}\left( \tilde\nabla_s \tilde\nabla_m g_{rn} + \tilde\nabla_s \tilde\nabla_n g_{rm} \right) \nl + \frac{1}{2} \tilde g^{rs}\tilde\nabla_m \tilde\nabla_n g_{rs} +  \frac{1}{2} \tilde\nabla^2 g_{mn}  - \frac{1}{2}  \left(\tilde\Pi_{mn}-\frac{D-n}{D-2}\tilde g_{mn}\right) g_s \left(\frac{1}{\tilde V}-\tilde\delta(\Sigma)\right),\\
0 &= - \d F_{n-1} \mp (-1)^{n(n-1)/2}\left(\frac{1}{\tilde V}-\tilde\delta(\Sigma)\right) \tilde{\text{dvol}}_{n}. \label{e4}
\end{align}
Note that the right-hand sides of all equations consistently integrate to zero on a compact space. An explicit step-by-step derivation of the above equations for the DGKT example can be found in \cite{Junghans:2020acz}.

It is straightforward to verify that \eqref{e1}--\eqref{e4} are solved by
\begin{align}
\e^{A} &= 1 + \frac{1}{4} g_s G, \label{a1} \\
g_{mn\parallel} &= \tilde g_{mn\parallel} \left(1 + \frac{1}{2} g_sG \right) \label{a2}, \\
g_{mn\perp} &= \tilde g_{mn\perp} \left(1 - \frac{1}{2} g_sG \right) \label{a3}, \\
\e^{-\phi} &= g_s^{-1} \left(1 + \frac{2n-D-2}{8} g_s G \right), \label{a4} \\
F_{n-1} &= \tilde F_{n-1} \pm (-1)^{n(n+1)/2} \tilde\star_n\d G. \label{a5}
\end{align}
Here
$g_{mn\parallel}$, $g_{mn\perp}$ are the parallel and transverse internal metric components, respectively.
We furthermore denote by $G$ the Green's function of the Laplacian on the transverse space, which satisfies
\begin{equation}
\tilde\nabla^2 G = \frac{1}{\tilde V} - \tilde\delta(\Sigma). \label{poisson}
\end{equation}
The corresponding expressions for an (anti-)D-brane would be obtained by flipping $G\to -G$ in \eqref{a1}--\eqref{a5}.
It is also straightforward to generalize the above solution to the case of several sources. The leading correction to each field is then simply the sum over the corrections for the individual sources.

Above we only displayed the leading corrections for the fields which are directly sourced by the O-plane. Some of the other fields can in general also receive corrections (in particular, the exact piece of $F_{n-1}$, other form fields like $F_{n-3}$ and $H_3$ or some of the metric components that vanish in the smeared solution), which are obtained in terms of the displayed ones using the linearized equations of motion.
To demonstrate this, we state here the leading corrections to $F_{n-3}$ and $H_3$ for the simplest case where  no further fluxes are present in the smeared solution (i.e., $\tilde F_q=0$ for $q\neq n-3$). More complicated setups work analogously by solving the corresponding linear equations. Let us also assume that the $n$-dimensional transverse space is a product of an $(n-3)$-dimensional space and a 3-dimensional one on which $\tilde F_{n-3}$ and $\tilde H_3$ are harmonic forms, respectively. The relevant equations are then \eqref{eom4} and \eqref{eom5}, which at linear order in the backreaction become
\begin{align}
\d \left( \tilde \star_n F_{n-3} \right) &= \mp (-1)^{n(n-1)/2}\d G \w \tilde H_3 - \frac{D-6}{4}g_s \d G \w \tilde\star_n \tilde F_{n-3}, & \d F_{n-3} &= 0, \\
\d \left( \tilde \star_n H_3 \right) &= \pm (-1)^{n(n+1)/2} g_s^2 \d G \w \tilde F_{n-3} - g_s \d G \w \tilde\star_n \tilde H_3, & \d H_3 &= 0,
\end{align}
where we used \eqref{a1}--\eqref{a5}.
This is solved by\footnote{To see this, note that $\d  \left[\tilde \star_n \d \left( \alpha \tilde H_3\right)\right] = (-1)^n(g_sG+\text{const.})\,\tilde\star_n \tilde H_3 +\tilde \star_n \d \left[\tilde \star_n \d \left( \alpha \tilde\star_n\tilde H_3\right) \right]$ and analogously for the other exact terms.}
\begin{align}
F_{n-3} &= \tilde F_{n-3} \pm (-1)^{n(n-1)/2}\frac{1}{g_s}\d \left[ \tilde \star_n \d \left( \alpha \tilde H_3\right)\right] +\frac{D-6}{4} \d \left[\tilde \star_n \d \left( \alpha \tilde\star_n \tilde F_{n-3}\right)\right], \label{a6} \\
H_3 &=\tilde H_3 \mp (-1)^{n(n-1)/2} g_s \d \left[\tilde \star_n \d \left( \alpha \tilde F_{n-3}\right)\right] +(-1)^n \d \left[ \tilde \star_n \d \left( \alpha \tilde\star_n \tilde H_3\right)\right], \label{a7}
\end{align}
where the function $\alpha$ satisfies $\tilde\nabla^2 \alpha = g_sG+\text{const.}$~and the constant is chosen such that integrating the right-hand side over the transverse space gives zero.

\subsubsection{Validity of the expansion and relation to singularities}
\label{sec:val}

Now that we know the leading backreaction corrections to the smeared solution, we need to determine when they are small.
One might naively guess that the regime of small backreaction is simply $g_sG \ll 1$. However, we claim that
the parametrically stronger condition
\begin{equation}
D g_s G \ll 1 \label{smallb}
\end{equation}
is required up to $\mathcal{O}(1)$ factors at large $D\gg 10$.
To see this, consider the situation where the RR tadpole of the O-plane is cancelled by a flux $\int H_3\w F_{n-3}$. Using the solution of Section \ref{sec:single}, we find that the energy density of the RR flux is
\begin{equation}
\e^{2\phi}|F_{n-3}|^2 = g_s^2 \left(1+\frac{D-4}{4}g_sG\right) |\tilde F_{n-3}|^2 + 2 \tilde\nabla_{n-3}^2\alpha \frac{g_s}{\tilde V} - \frac{D-6}{2} \tilde\nabla_{n-3}^2\alpha |g_s\tilde F_{n-3} |^2 \label{fn0}
\end{equation}
at linear order in the backreaction, where the tildes on the right-hand side indicate as before that the objects are constructed with the smeared fields. Recall further that $\alpha$ satisfies $\tilde\nabla^2\alpha=g_sG+\text{const.}$, where $\tilde\nabla^2=\tilde\nabla_{n-3}^2+\tilde\nabla_3^2$ is the Laplacian on the $n$-dimensional space transverse to the O-plane. By $\tilde\nabla_{n-3}^2$ and $\tilde\nabla_{3}^2$, we mean the Laplacians on the $(n-3)$-dimensional and $3$-dimensional subspaces on which $\tilde F_{n-3}$ and $\tilde H_3$ live, respectively.

We observe that the energy density of the RR flux is corrected by three terms: one term $\sim Dg_sG$ and two terms $\sim \tilde\nabla_{n-3}^2\alpha$. Importantly, the $Dg_sG$ term does not cancel with the $\tilde\nabla_{n-3}^2\alpha$ terms at a generic point on the transverse space. Indeed, for $n=3$, the latter just vanish. For $n>3$, they do not vanish in general (except in special solutions like GKP \cite{Giddings:2001yu}) but are different functions than the $Dg_sG$ term and thus cannot cancel the latter at a generic point. We thus conclude that we should demand \eqref{smallb} to control the backreaction corrections, as claimed above.\footnote{A subtlety occurs in the case where $n$ is large, but this does not change the conclusion.
In particular, let us assume a generic point on the transverse space and no parametrically large anisotropies so that derivatives of $\alpha$ along the 3-dimensional subspace with $\tilde H_3$ flux are not parametrically larger than derivatives of $\alpha$ along the remaining $n-3$ dimensions. We then expect that $\tilde\nabla^2_{3}\alpha \ll \tilde\nabla^2_{n-3}\alpha$ at large $n$ and therefore $\tilde\nabla^2_{n-3}\alpha \to \tilde\nabla^2\alpha=g_sG$ up to $1/n$ corrections. 
The $\tilde\nabla^2_{n-3}\alpha$ corrections and the $Dg_sG$ corrections in \eqref{fn0} might thus cancel each other at leading order, which would invalidate our estimate for the backreaction. However, we find $\e^{2\phi}|F_{n-3}|^2 = g_s^2 \left(1-\frac{D}{4}g_sG + \mathcal{O}(\frac{D}{n}g_sG)\right) |\tilde F_{n-3}|^2 + 2 \left(g_sG+ \mathcal{O}(\frac{1}{n}g_sG)\right) \frac{g_s}{\tilde V} $, where the $\frac{g_s}{\tilde V}$ term can be neglected since $\frac{g_s}{\tilde V}\lesssim g_s^2|\tilde F_{n-3}|^2$ in solutions with O-planes (cf.~the discussion around \eqref{dilstab}). The leading corrections to the RR energy density are thus $\sim Dg_sG$ so that we again require \eqref{smallb} just as for small $n$.}

In models where the RR tadpole of the O-plane is not cancelled by fluxes but by an anti-O-plane, deriving the condition \eqref{smallb} is a bit more subtle.\footnote{Our discussion also applies if the tadpole is cancelled by D-branes but this cannot happen for all O-plane charges in the dS case since a net negative tension must remain according to \eqref{mn}.}
One way to motivate it is to consider the backreaction of the O-plane on the anti-O-plane (or vice versa) and demand that this should be small.
The action of the anti-O-plane in the background generated by the O-plane is
\begin{equation}
\frac{2\pi}{g_s} \int \d^{D-n} x \sqrt{-\tilde g_{D-n}} \left(1 + \frac{D-2}{8}g_sG\right) \label{ao0}
\end{equation}
at linear order in the backreaction, where $G$ denotes the Green's function associated to the O-plane (i.e., $\tilde\nabla^2 G = \frac{1}{\tilde V} - \tilde\delta(\Sigma)$, where $\tilde\delta(\Sigma)$ has support on the O-plane, not the anti-O-plane).
We thus again observe the crucial $D$ factor in the integrand, leading to \eqref{smallb}.\footnote{Since \eqref{poisson} only determines $G$ up to a constant, one might wonder whether this freedom can be used to set the corrections at the locus of the anti-O-plane to zero. However, once the moduli are fixed, it is clear that there cannot be any free parameters. The constant in $G$ must therefore be fixed by the equations of motion at higher orders in the backreaction expansion.}
One might complain here that one should not expect our expansion to be valid at the source locations in the first place. Indeed, the backreaction of a localized source diverges in its own vicinity and we should not plug this back into its own action. However, we believe it still makes sense to demand as above that the backreaction of a source is small at the location of the \emph{other} sources since otherwise the smeared approximation stops being reliable.

An alternative argument is that corrections to the transverse metric determinant $\sqrt{g_n}=\sqrt{\tilde g_n}\left(1-\frac{n}{4}g_sG\right)$ should be small since the transverse volume appears in the definition of the smeared source terms. This then again implies \eqref{smallb} unless $n\ll \mathcal{O}(D)$ in which case \eqref{smallb} follows from demanding small corrections in \eqref{a4}.
Finally, we expect that the condition \eqref{smallb} may also follow from various higher-than-linear corrections in the equations of motion, which we do not compute here.

\bigskip

We have seen that the O-plane backreaction is small at points on the manifold where \eqref{smallb} holds. At all other points, the backreaction is non-linear and the expansion around the smeared solution breaks down. For finite $g_s$, this inevitably happens in a finite region surrounding the O-plane since $G$ diverges like $1/r^{n-2}$ (for $n>2$) at small distances $r$. As explained before, we furthermore expect a singular hole to open up in this non-linear region.
In the limit $D g_s G\to 0$, the size of this singular hole goes to zero and the smeared approximation becomes reliable everywhere on the manifold, see Fig.~\ref{smeared}.

Let us illustrate this in an explicit example. We consider an O3-plane in GKP, i.e., a Calabi-Yau orientifold of type IIB string theory with $D=10$, $d=4$, $n=6$. The solution including the full non-linear backreaction is known in this setup and given by $g_{mn}=\e^{-2A}\tilde g_{mn}$, $\e^{-4A}=1-g_sG$, $\e^\phi=g_s$ \cite{Giddings:2001yu}. This yields
\begin{equation}
\e^{2\phi}|F_{3}|^2 = \frac{g_s^2|\tilde F_{3}|^2}{\left(1 - g_s G\right)^{3/2}}. \label{fgkp}
\end{equation}
Using that the smeared GKP solution satisfies $g_s^2|\tilde F_3|^2=|\tilde H_3 |^2=\frac{g_s}{\tilde V}$ \cite{Blaback:2010sj}, one can check that \eqref{fgkp} agrees with \eqref{fn0} at linear order in $g_sG$ as it should.

What happens in the non-linear regime?
As we move from a region of small backreaction
towards the O3-plane, $g_sG$ becomes larger in the denominator of \eqref{fgkp} until
we eventually reach a singularity at $g_sG=1$.
At all points with $g_sG > 1$, the energy density is formally imaginary, which is incompatible with a positive dilaton and a positive-definite metric and clearly indicates that the classical solution is no longer trustworthy.
It is natural to expect that string corrections blow up in this region and significantly modify the classical solution to cure the naive singularity.
We thus have large string corrections for $g_sG \gtrsim 1$, which is the same regime where the backreaction becomes non-linear.

To summarize, O-planes are surrounded by a stringy region where the classical approximation breaks down. From the point of view of the (backreacted) classical solution, this region looks like a singular hole. Its size can be estimated by determining the locus in spacetime where the backreaction is non-linear.

Returning to general $D\gg 10$, we expect that the classical solution breaks down for $g_sG\gtrsim 1/D$ since as we showed the backreaction becomes non-linear there.
Strictly speaking, this is an assumption since
neither the full non-linear backreaction nor the stringy resolution of the classical singularities are known in general compactifications (in particular, in the supercritical case). We can therefore not exclude that there are backgrounds where for some reason O-planes backreact more mildly than in the known examples (see App.~\ref{app:caveat} for a toy model). However, in all examples we are aware of, finite-distance singularities are indeed a characteristic feature of an O-plane in the regime where the non-linear backreaction sets in. Moreover, it follows from
\eqref{fn0}, \eqref{ao0} that, even in the absence of stringy effects, the classical backreaction alone would already significantly modify the smeared solution for $g_sG\gtrsim 1/D$. We therefore in any case lose control in this regime unless we compute the solution non-perturbatively in $g_sG$, which would be very difficult in practice.
For these reasons, we believe that \eqref{smallb} is a well-motivated condition for control, and we will impose it in the remainder of this paper.

\subsubsection{Backreaction of several sources}

How does the condition $Dg_sG\ll 1$ generalize to the case of several sources? We claim that the appropriate generalization is
\begin{equation}
Dg_s \sum_i G_i \ll 1, \label{gb0}
\end{equation}
where $i$ runs over all sources and $G_i$ are the corresponding Green's functions. It is straightforward to verify this if all sources are wrapped on the same cycle. Indeed, the total correction to a field due to the backreaction of all sources is at the linear level simply the sum over the corrections due to the individual sources. Therefore, all we have to do is repeat the argument of Section \ref{sec:val} replacing $G$ by $\sum_i G_i$ in all expressions, which yields \eqref{gb0}.

On the other hand, if the sources wrap different, possibly intersecting cycles $\mathcal{C}_a$, it is less obvious that \eqref{gb0} is the correct condition since the Green's functions then do not just appear in the corrections as the sum $\sum_i G_i$ but with different prefactors that depend on the codimensions of the sources and on how they overlap. For example, in a metric component parallel to sources on $\mathcal{C}_1$ and transverse to sources on $\mathcal{C}_2$, the Green's functions of the two sets of sources have a relative minus sign: $g_{mn} = \tilde g_{mn} \left(1 + \frac{1}{2} g_s\sum_{i\in \mathcal{C}_1} G_i -\frac{1}{2} g_s \sum_{i\in \mathcal{C}_2} G_i \right)$ (cf.~\eqref{a2}, \eqref{a3}).
Similarly, the two sums of Green's functions can appear with relative prefactors in the corrections to the RR energy densities. It is therefore not straightforward to repeat the argument of Section \ref{sec:val} in this case.
In particular, one may wonder whether there could be cancellations between corrections from $\mathcal{C}_1$ and $\mathcal{C}_2$ so that the total backreaction is small even when \eqref{gb0} is parametrically violated.
However, this cannot happen at generic points since, by assumption, the Green's functions of the sources on $\mathcal{C}_1$ and $\mathcal{C}_2$ depend on different coordinates so that they can at best cancel at a measure-zero subset of the points. We should therefore demand that the corrections associated to each cycle are small \emph{individually}, i.e., $Dg_s \sum_{i\in \mathcal{C}_a} G_i \ll 1$ for each $\mathcal{C}_a$ on which sources are wrapped. Summing this over $a$ then again implies \eqref{gb0} up to $\mathcal{O}(1)$ factors as in the single-cycle case.

This argument assumes that the number of cycles with wrapped sources is $\mathcal{O}(1)$ but leads to a parametrically weaker condition than \eqref{gb0} if we put sources on parametrically many (e.g., $\mathcal{O}(D)$) different cycles. We will not present a fully general proof of \eqref{gb0} in this latter case. However, such tricks to avoid parametric problems do typically not work in string theory, and we anticipate that the naively gained factor $D$ may be lost again elsewhere so that one again arrives at \eqref{gb0}.

In Section \ref{sec:b3}, we will discuss a class of models with $H_3$ and $F_{n-3}$ fluxes where it is not difficult to derive \eqref{gb0}
although the O-planes wrap $\mathcal{O}(D)$ different cycles. The reason is that the O-planes in these models are oriented such that $F_{n-3}$ is transverse to all of them. The expression for the RR energy density is thus
\begin{equation}
\e^{2\phi}|F_{n-3}|^2 = g_s^2\left( 1 + \frac{D-4}{4}g_s \sum_i G_i \right)|\tilde F_{n-3}|^2 + \ldots,
\end{equation}
where $i$ runs over all sources on all cycles. The dots stand for additional terms with a different functional dependence (like the $\tilde\nabla_{n-3}^2\alpha$ terms in \eqref{fn0}). We thus find that controlling the backreaction requires \eqref{gb0} in these models in spite of the parametrically large number of wrapped cycles.

Let us finally also reinstate the dependence on the tension, which we have set to $\frac{T_{p_i}}{2\pi}=-1$ until now. The condition for small backreaction is thus
\begin{equation}
D g_s\sum_i |T_{p_i}| G_i \ll 1, \label{gb}
\end{equation}
where we in general allow sources with negative and positive tension.
Note that the absolute value $|T_{p_i}|$ appears instead of $T_{p_i}$ here since
the combined backreaction of two sources with opposite tensions $T_{p_1}=-T_{p_2}$ is generically of the order $T_{p_1}$ rather than cancelling out (recall that we do not care about a factor 2 but only about the parametric behavior). This is true unless these sources are placed on top of each other (or at a very small distance), which would be captured by simply setting $T_{p_i}=0$ for $i=1,2$ in the above sum.
For dS vacua, it cannot happen that $T_{p_i}=0$ for all $i$ since a net negative tension must remain to satisfy \eqref{mn}.
It is therefore not possible to trivially satisfy \eqref{gb} by locally cancelling all O-plane tensions using D-branes.

\subsubsection{The small-hole condition}
\label{sec:shc}

Where on the manifold should we impose \eqref{gb}? Since $G_i$ diverges like $1/r^{n-2}$ at small distances $r$ from a source, it is inevitable that the backreaction becomes large sufficiently close to it (cf.~Fig.~\ref{smeared}). It is therefore clear that \eqref{gb} cannot hold \emph{everywhere} (assuming that $g_s$ is finite).
One might now give up and claim that compactifications with O-planes cannot be described classically and instead always require a fully non-perturbative computation. However, we believe this is too pessimistic. A milder condition proposed in \cite{Cribiori:2019clo, Junghans:2020acz} (see also \cite{Gao:2020xqh}) is the following:
\begin{quote}
{\bf Small-Hole Condition (SHC):} \emph{The classical description of a string compactification with O-planes is reliable if the singular/stringy regions generated by the O-planes cover a sufficiently small fraction of the compact space.}
\end{quote}
We stress that this is obviously only a necessary and not a sufficient condition for a classical solution to be a consistent string vacuum, as there might be additional problems that are not visible at the level of the classical equations of motion but only at higher orders in the $\alpha^\prime$ expansion or non-perturbatively.

How small is ``sufficiently small''? The idea behind the SHC is that the classical description is reliable if
the local divergences near the O-planes, and the associated stringy effects, do not significantly affect the long-distance physics.
In particular, we should demand that $d$-dimensional observables (such as moduli masses, couplings or the cosmological constant) can reliably be computed using the classical solution, in spite of its breakdown at very short distances.
This should precisely be the case when the regions where the backreaction blows up are small enough
that their contribution to integrals over the compact space is negligible. Technically, this is motivated by the fact that the $d$-dimensional effective action is obtained by dimensional reduction of the $D$-dimensional one, i.e., by integrating the $D$-dimensional Lagrangian over the compact space.

The prime example illustrating the logic behind the SHC is again DGKT, where the fraction of the compact space covered by the stringy regions near the O6-planes goes to zero in the large-flux limit $N\to\infty$ (measured in the smeared metric). Proving that string corrections to $d$-dimensional observables become negligible in this limit is of course hard without explicitly knowing the full non-perturbative solution. However, a parametric estimate suggests that, for large enough $N$, the 4d scalar potential is indeed well-approximated by the smeared solution in spite of the local divergences near the O6-planes \cite[Sec.~5]{Junghans:2020acz}.

It is natural to wonder whether large $D$ could in a similar way serve as a control parameter for supercritical dS models, in the sense that the stringy regions surrounding the O-planes shrink to zero in the limit $D\to\infty$. However, this is not the case, as we will see in the remainder of this paper. In particular, we will argue that the SHC is violated in supercritical dS models.

\subsection{Estimating the Green's function}
\label{sec:b2}

We argued in Section \ref{sec:b1} that the condition $Dg_s \sum_i|T_{p_i}| G_i \ll 1$ is a reasonable proxy for small backreaction and control over the classical solution. We would now like to understand how this condition constrains the geometry of the spaces transverse to the sources. To this end, we have to estimate the Green's functions.

Let us again consider a single source of codimension $n= D-p-1$ and an associated Green's function $G$.
We choose to normalize $G$ such that $G=0$ at its minimum, which can always be done by a constant shift.\footnote{Note that, according to \eqref{poisson}, our sign convention for $G$ is such that it grows approaching the source and consequently it must have a minimum somewhere in the compact space.}
To simplify the notation, we will furthermore drop the tildes on top of Laplacians, volumes, etc.~in the following. It is understood that these quantities are always defined using the smeared metric.

How can we estimate the magnitude of $G$ on a general $n$-dimensional manifold? A quick estimate is to consider the Green's function in flat space $G\sim r^{-(n-2)}$ (for $n\neq 2$), where $r$ is the radial distance to the source. Evaluating this at a generic distance $r\sim \mathcal{O}(R)$, where $R$ is a length scale characterizing the compact space, yields
\begin{equation}
G \sim \frac{1}{R^{n-2}}. \label{dgsagf}
\end{equation}
The same conclusion is reached by considering the defining equation
\begin{equation}
\nabla^2 G = \frac{1}{V}-\delta(\Sigma), \label{gdef}
\end{equation}
where $V$ denotes the volume. Since the Laplacian scales like $R^{-2}$ and the volume scales like $R^{n}$,
we conclude again that $G$ must scale as in \eqref{dgsagf}, without having to assume anything specific on the manifold.
A familiar example is the warping generated by a D3-brane on a Calabi-Yau 3-fold in type IIB string theory. Since the D3-brane is pointlike on the Calabi-Yau, we have $n=6$. The length scale $R$ is in this case determined by the string-frame Calabi-Yau volume as $\mathcal{V} \sim R^6$. The warping corrections are thus of the order $g_sG \sim \frac{g_s}{R^4}$ at generic points on the Calabi-Yau. At very small distances $r \ll R$ from the source, the backreaction is of course much stronger, as $G$ then diverges like $1/r^4$.

The estimate \eqref{dgsagf} is often very helpful to get an intuition for the backreaction. Indeed, it correctly predicts how the backreaction behaves under rescalings of the overall volume while keeping the shape of the geometry fixed. In particular, we can read off from \eqref{dgsagf} that the backreaction becomes small in the large-volume regime (for $n\ge 3$). Nevertheless, relying on this estimate can be deceptive for (at least) two reasons, as we now explain.

\subsubsection{Anisotropy}
\label{sec:anisotr}

The first reason is that the space can be highly anisotropic, i.e., be characterized by several, parametrically different length scales. As a simple example, consider a 3-torus
consisting of one large circle with radius $R_1$ and two much smaller ones with radii $R_2,R_3\ll R_1$. We choose a metric $\d s^2 = \sum_{i=1}^3(2\pi R_i)^2 \d x_i^2$ with $x_i \in [0,1]$ and put the source at $x_i=0$. We now ask how the Green's function $G_{\mathbb{T}^3}(\vec x)$ of the 3d Laplacian scales with the radii on such a space. We expect that, at all points with
$x_1 \gg \frac{R_2}{R_1},\frac{R_3}{R_1}$, the Green's function looks effectively like the Green's function on a circle. At large $R_1$, this is true almost everywhere on the torus. See \cite[App.~D]{Cribiori:2021djm} for a numerical verification of this claim (and an analytic proof for the simpler example of the 2d cylinder). We can thus make the ansatz $G_{\mathbb{T}^3}(\vec x) = C (x_1^2-|x_1|+\frac{1}{4})$ up to small corrections. Here $C$ is a constant which can be fixed by demanding that we recover the ordinary circle Green's function $G_{S^1}(x_1)= \pi R_1 (x_1^2-|x_1|+\frac{1}{4})$ when integrating over the two small dimensions: $\int \d x_2 \d x_3 \sqrt{g}\, G_{\mathbb{T}^3}(\vec x) =  \pi R_1 (x_1^2-|x_1|+\frac{1}{4})$. This yields
\begin{equation}
G_{\mathbb{T}^3}(\vec x) = \frac{R_1}{4\pi R_2R_3} \left(x_1^2-|x_1|+\frac{1}{4}\right), \qquad x_1 \gg \frac{R_2}{R_1},\frac{R_3}{R_1}.
\end{equation}
At a generic point on the torus, the parametric scaling of $G_{\mathbb{T}^3}$ is thus $G_{\mathbb{T}^3}\sim \frac{R_1}{R_2R_3}$. For the case $R_i=R$, this would reduce to $G_{\mathbb{T}^3}\sim \frac{1}{R}$, in agreement with \eqref{dgsagf} for $n=3$. However, making $R_1$ large at fixed volume, the backreaction becomes much stronger.

It is useful to write the scaling of the Green's function in a unified way by expressing it in terms of the volume $V=8\pi^3R_1R_2R_3$ and the diameter $\mathcal{D}\approx \pi R_1$:
\begin{equation}
G \sim \frac{\mathcal{D}^2}{V}. \label{est1}
\end{equation}
This correctly reproduces the scaling for the isotropic case $R_i=R$ and in all anisotropic limits where one of the radii is much larger than the other two.

In fact, we claim that \eqref{est1} is a reliable estimate not only on the torus but on a general manifold (aside from a dimension-dependent factor determined further below). First of all, the inverse volume scaling in \eqref{est1} follows immediately from \eqref{gdef} at any point away from the source. For dimensional reasons, we have to multiply this with a squared length scale. As the 3-torus example demonstrates, this should be the longest length scale available, so it is natural to multiply by $\mathcal{D}^2$.
In particular, in a regime where a manifold is stretched along $n_\text{eff}$ of its $n$ dimensions in such a way that it becomes very long and thin, it effectively looks $n_\text{eff}$-dimensional at all distances from the source which are larger than the size of the small dimensions. The Green's function in this regime should then agree with the corresponding $n_\text{eff}$-dimensional Green's function at generic points on the manifold and therefore scale with the diameter like $G =G_0 \mathcal{D}^{2-n_\text{eff}}$, where $G_0$ is a factor which does not depend on $\mathcal{D}$. This is precisely reproduced by \eqref{est1} since $V=V_0\mathcal{D}^{n_\text{eff}}$ in the stretched regime.

\subsubsection{Large dimension}

A second caveat with \eqref{dgsagf} is that, in the large-$n$ regime, we have to be very careful with $n$-dependent factors since these can parametrically affect the scaling of $G$. This is particularly relevant for this work: since we are interested in string compactifications at large $D$, also $n$ can be large.

As an example, consider the Green's function of the $n$-sphere with radius $R$:
\begin{equation}
G_{S^n}(\theta) = \frac{\mathcal{D}^2}{V} \left( \frac{1}{\pi^{2}}\int\d \theta \left[ \sin(\theta)^{1-n} \left(\int \d \theta \sin(\theta)^{n-1}+C_1\right) \right] + C_2\right), \label{gsphere}
\end{equation}
where $\theta=[0,\pi]$ is the polar angle and $\mathcal{D}=R \pi$ is the diameter. This can be verified by integrating \eqref{gdef} using the standard Laplacian in spherical coordinates. The integrals are indefinite ones, where we absorb the arbitrary integration constants into $C_1$ and $C_2$.
We choose the integration constant $C_1$ such that the source sits at $\theta=0$ and $G_{S^n}^\prime=0$ at $\theta=\pi$. As stated before, we choose the second constant $C_2$ such that $\text{min}(G_{S^n})=G_{S^n}(\pi)=0$.

We first observe that, at large $n$, the estimate \eqref{est1} is super-exponentially larger than the estimate \eqref{dgsagf}:
\begin{equation}
\frac{\mathcal{D}^2}{V} = \frac{R^{2-n} \pi^2 \Gamma\left(\frac{n+1}{2}\right)}{2\pi^{\frac{n+1}{2}}} \simeq \frac{\pi^2}{\sqrt{2}}\left(\frac{n}{2\pi \e}\right)^{n/2}R^{2-n}.
\end{equation}
This demonstates that, at large $n$, we need to carefully distinguish length scales even in an isotropic space.
In particular, the two length scales $V^{1/n}$ and $R$ differ parametrically at large $n$ for the $n$-sphere, leading to the large discrepancy between \eqref{est1} and \eqref{dgsagf}.

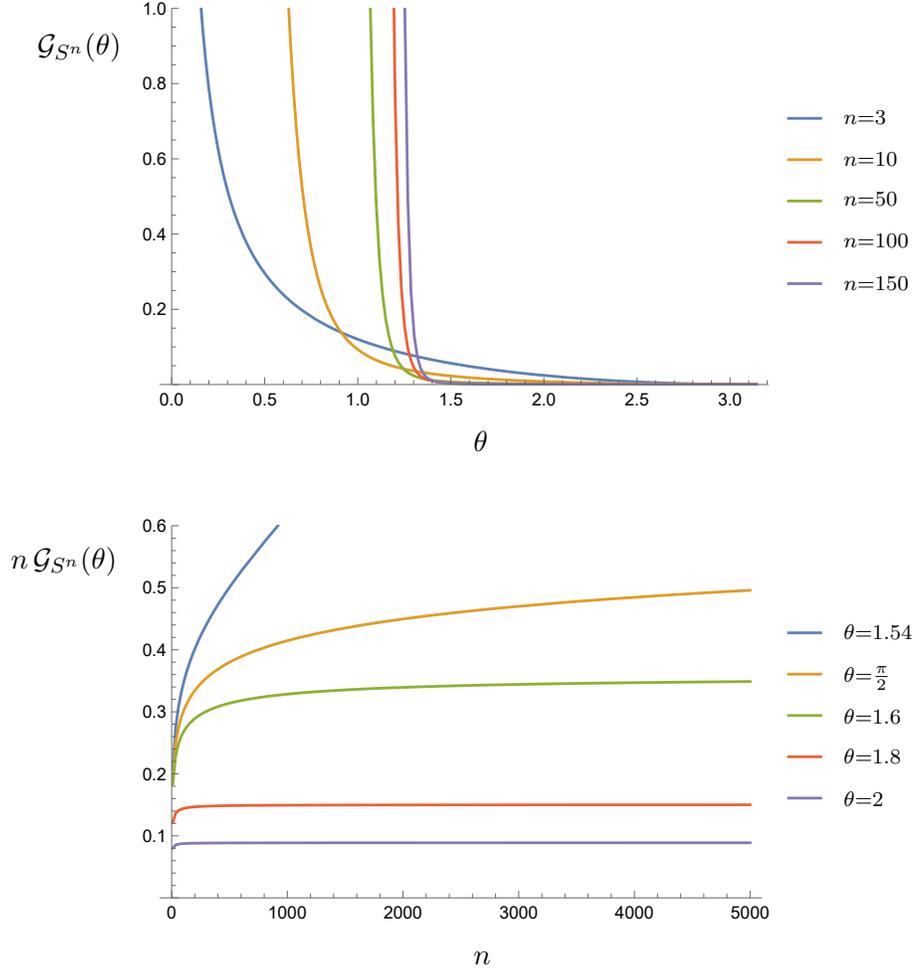
\begin{figure}[t!]
\centering
\includegraphics[trim = 0mm 0mm 0mm 0mm, clip, width=0.6\textwidth]{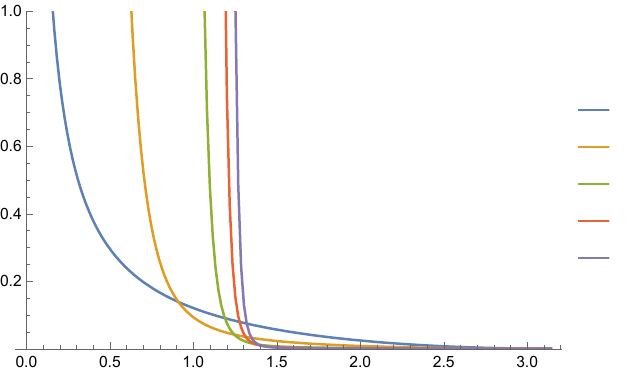}
\put(-140,-15){${\theta}$}
\put(-305,135){$\mathcal{G}_{S^n}(\theta)$}
\put(0,109){$\scriptstyle{n=3}$}
\put(0,93){$\scriptstyle{n=10}$}
\put(0,78){$\scriptstyle{n=50}$}
\put(0,62){$\scriptstyle{n=100}$}
\put(0,46){$\scriptstyle{n=150}$}
\\[2em]

\includegraphics[trim = 0mm 0mm 0mm 0mm, clip, width=0.6\textwidth]{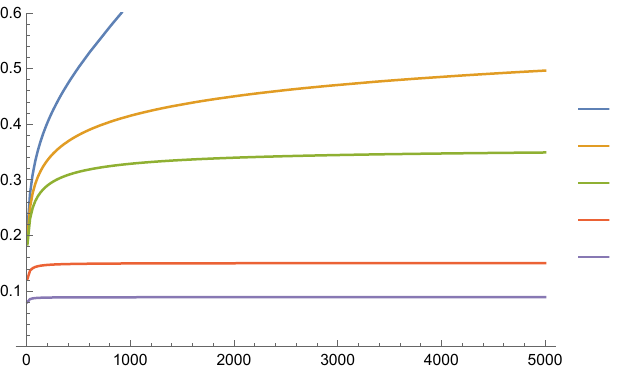}
\put(-140,-15){${n}$}
\put(-315,135){$n\,\mathcal{G}_{S^n}(\theta)$}
\put(0,108){$\scriptstyle{\theta=1.54}$}
\put(0,92){$\scriptstyle{\theta=\frac{\pi}{2}}$}
\put(0,76){$\scriptstyle{\theta=1.6}$}
\put(0,61){$\scriptstyle{\theta=1.8}$}
\put(0,45){$\scriptstyle{\theta=2}$}
\caption{ Profile of the Green's function of the $n$-sphere at fixed $\frac{\mathcal{D}^2}{V}=1$ for different choices of $n$ (top). At large $n$, the Green's function scales like $\gtrsim 1/n$ (bottom).
\label{nsphere}}
\end{figure}

The overall factor $\frac{\mathcal{D}^2}{V}$ in \eqref{gsphere} appears to support the proposal \eqref{est1}. However, to verify this, we have to study the $\theta$-dependent factor inside the brackets in \eqref{gsphere}, which we denote as $\mathcal{G}_{S^n}\equiv \frac{V}{\mathcal{D}^2}G_{S^n}$ from now on.
For $n=\mathcal{O}(1)$, we find that $\mathcal{G}_{S^n}$ is an $\mathcal{O}(1)$ function of $\theta$, except at small $\theta$ where it diverges, see Fig.~\ref{nsphere}. We thus obtain $G_{S^n}\gtrsim \frac{\mathcal{D}^2}{V}$, in agreement with \eqref{est1}.

At large $n$, things are more subtle. Taking the limit of large $n$ (at fixed $\frac{\mathcal{D}^2}{V}$), we observe that the source becomes very thick, with a hard wall at $\theta=\theta_*\approx \frac{\pi}{2}$ separating a region of vanishing backreaction from a region of huge backreaction, see Fig.~\ref{nsphere}.
Our numerical analysis suggests that the wall approaches $\theta_*\to \frac{\pi}{2}$ in the limit $n\to\infty$. At large $n$, it is therefore not accurate to say that $\mathcal{G}_{S^n}$ is an $\mathcal{O}(1)$ function. Rather, $\mathcal{G}_{S^n}\to\infty$ in roughly one half of the $\theta$ interval and $\mathcal{G}_{S^n}\to 0$ in the other half.
One might be tempted to conclude from this that the region of large $\mathcal{G}_{S^n}$ occupies about half of the \emph{volume} of the sphere. However, as stated before, one should not jump to conclusions at large $n$. Indeed, we numerically verified that the fraction of the volume occupied by points with large $\mathcal{G}_{S^n}$ tends to \emph{zero} at large $n$, $\lim\limits_{n\to\infty} \frac{\int_0^{\theta_*(n)}\d\theta \sin^{n-1}(\theta)}{\int_0^{\pi}\d\theta \sin^{n-1}(\theta)}= 0$.

As we explained in Section \ref{sec:shc}, we believe it would be too strict to demand small backreaction \emph{everywhere} on the sphere. Instead, we would like to allow large backreaction in a small fraction of the volume and only demand small backreaction at \emph{generic} points. Let us therefore ignore the region $\theta\le \theta_*$ and focus on the region $\theta>\theta_*$ where most points on the $S^n$ lie. Our numerical analysis shows that $\mathcal{G}_{S^n}$ scales like $1/n$ in this region, see Fig.~\ref{nsphere}.
The $n$-sphere example thus suggests the following refinement of \eqref{est1}:
\begin{equation}
G \gtrsim c_G(n)\frac{\mathcal{D}^2}{V}, \qquad c_G(n) \sim \left\{\begin{array}{lll}
                                                            \displaystyle{1} & \qquad & n=\mathcal{O}(1) \\[0.5em] \displaystyle{\frac{1}{n}} & \qquad & n\gg 1
                                                            \end{array}\right.. \label{est2}
\end{equation}
An obvious exception is the minimum of $G$ where in our conventions $G=0$.

\begin{figure}[t!]
\centering
\includegraphics[trim = 0mm 0mm 0mm 0mm, clip, width=0.6\textwidth]{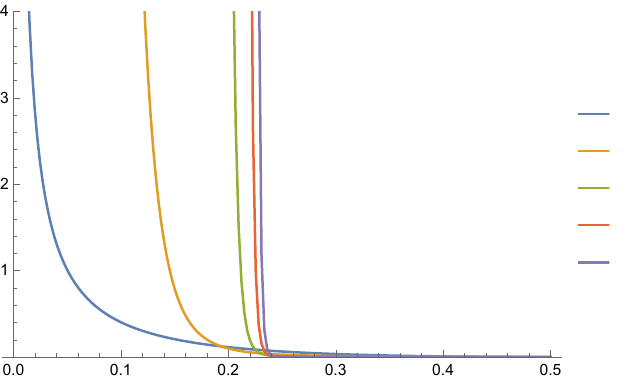}
\put(-140,-15){${x}$}
\put(-305,135){$\mathcal{G}_{\mathbb{T}^n}(x)$}
\put(0,109){$\scriptstyle{n=3}$}
\put(0,93){$\scriptstyle{n=10}$}
\put(0,78){$\scriptstyle{n=50}$}
\put(0,62){$\scriptstyle{n=100}$}
\put(0,46){$\scriptstyle{n=150}$}
\\[2em]

\includegraphics[trim = 0mm 0mm 0mm 0mm, clip, width=0.6\textwidth]{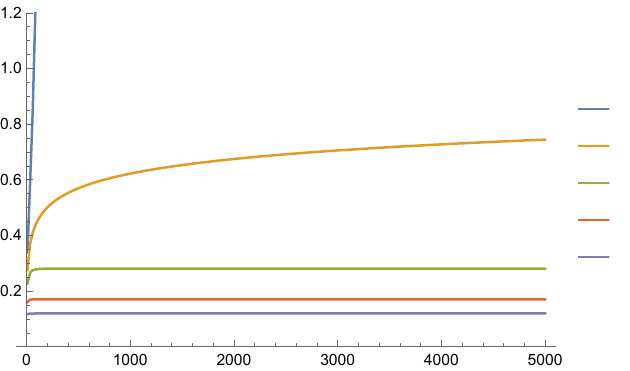}
\put(-140,-15){${n}$}
\put(-315,135){$n\,\mathcal{G}_{\mathbb{T}^n}(x)$}
\put(0,108){$\scriptstyle{y=0.24}$}
\put(0,92){$\scriptstyle{y=0.25}$}
\put(0,76){$\scriptstyle{y=0.26}$}
\put(0,61){$\scriptstyle{y=0.28}$}
\put(0,45){$\scriptstyle{y=0.3}$}
\caption{ Profile of the Green's function of the $n$-torus along the diagonal $x_i=x$ at fixed $\frac{\mathcal{D}^2}{V}=1$ for different choices of $n$ (top). At large $n$, the Green's function scales like $\gtrsim 1/n$ (bottom).
\label{ntorus}}
\end{figure}
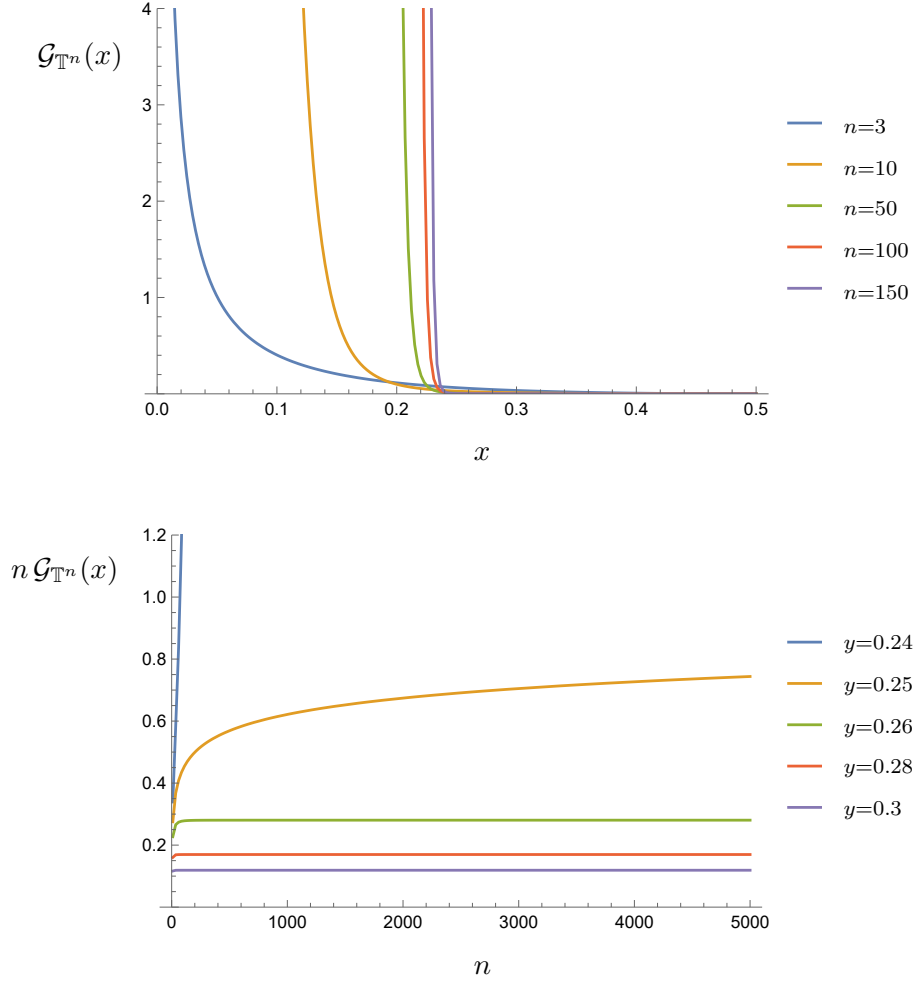

Let us study another example, the $n$-torus. For simplicity, we choose equal radii $R_i=R$ for all $i=1,\ldots,n$. We further choose the fundamental domain to be $x_i\in[-\frac{1}{2},\frac{1}{2}]$ and put the source at $x_i=0$. The Green's function is then \cite{CourantHilbert, Shandera:2003gx, Andriot:2019hay}
\begin{equation}
G_{\mathbb{T}^n}(x_i) = \frac{4\mathcal{D}^2}{nV} \int_0^\infty \d s \left(\prod_{i=1}^n\theta_3\left(x_i, \e^{-4\pi^2s}\right) -1\right) \label{tg}
\end{equation}
up to a constant shift,
where we used $\mathcal{D}=\pi \sqrt{n}R$, $V=(2\pi R)^n$ and $\theta_3$ is the Jacobi theta function defined by
\begin{equation}
\theta_3(b,\e^{-a}) = \sum_{n=-\infty}^{\infty} \e^{-an^2+2\pi ibn}.
\end{equation}
Confirming \eqref{est2} for the $n$-torus thus amounts to showing that \eqref{tg} scales at least like $1/n$ at large $n$ and fixed $\frac{\mathcal{D}^2}{V}$.
To verify this, we define
\begin{equation}
\mathcal{G}_{\mathbb{T}^n}(x) = \frac{4}{n}\int_0^\infty \d s \left(\theta_3\left(x, \e^{-4\pi^2s}\right)^n -1 \right),
\end{equation}
which probes the Green's function along the diagonal $x_i = x$. This should be sufficient for our purpose of confirming the scaling $\gtrsim 1/n$ since the diagonal includes the point $x_i=\frac{1}{2}$, which is farthest away from the source (and therefore the backreaction is the weakest).
Our numerical analysis shows that, analogously to the $n$-sphere, the source again becomes a hard wall at large $n$, i.e., $\mathcal{G}_{\mathbb{T}^n}$ diverges for a subset of the points and scales like $1/n$ for the remaining ones (see Fig.~\ref{ntorus}). Although we will not discuss it explicitly here, it is straightforward to verify analogous results also for points which do not lie on the diagonal.

It would be interesting to determine the precise boundaries of the divergent region and the fraction of the volume covered by it, as we did in the $n$-sphere example. However, since this is much more involved than in the $n$-sphere case, we will not attempt this here.
In any case, our results show that $\mathcal{G}_{\mathbb{T}^n}\gtrsim 1/n$ and therefore the Green's function scales like $G_{\mathbb{T}^n}\gtrsim \frac{\mathcal{D}^2}{nV}$, providing another non-trivial check of \eqref{est2}.

\bigskip

In conclusion, \eqref{est2} correctly predicts the scaling of $G$ in several non-trivial examples, in particular in examples with large anisotropies and in examples with large $n$.
Interestingly, both the $n$-sphere and the $n$-torus exhibit the same $1/n$ scaling in \eqref{est2}. This suggests that \eqref{est2} is independent of the topology and the curvature of the manifold and a general property of Green's functions at large $n$.
A heuristic argument is that, for unit diameter and in a locally orthonormal frame, $\mathcal{G}=\frac{V}{\mathcal{D}^2}G$ satisfies $\sum_{m=1}^n\partial_m^2\mathcal{G} =1$ at the minimum, and so we have $\partial_m^2\mathcal{G}\sim \frac{1}{n}$ for each direction. We thus expect that $\mathcal{G}$ scales like $\frac{1}{n}$ near the minimum and keeps growing towards the source position where it diverges. This suggests that the estimate \eqref{est2} holds in general, in particular also for negatively curved manifolds.\footnote{For metrics with large anisotropies, we further expect that $n$ is replaced by the number of large dimensions $n_\text{eff}$ along which $\mathcal{G}$ should dominantly vary, as discussed in Section \ref{sec:anisotr}.}
Nevertheless, given that we have not explictly tested \eqref{est2} on many different manifolds, we leave open the possibility that manifolds might exist in which the backreaction is weaker at large $n$.
Proving a rigorous lower bound on $G$ that confirms \eqref{est2} is an interesting geometry question which we leave for the future.

As a final comment, we note that \eqref{est2} is consistent with a rigorous \emph{upper} bound on $G$ due to Cheng and Li \cite{cheng1981} (see also \cite[Ch.~3]{siu1987}). Consider a compact $n$-dimensional Riemannian manifold $\mathcal{M}$ whose Ricci curvature is bounded from below such that $\mathcal{R}_{mn}u^mu^n\ge (n-1)K$ for some $K\in \mathbb{R}$ and all unit tangent vectors $u^m$.
Further define the shifted Green's function $G_\text{CL}=G - \frac{\int_\mathcal{M} G}{V}$ such that $\int_\mathcal{M} G_\text{CL} = 0$. Then $G_\text{CL}$ satisfies the bound
\begin{equation}
G_\text{CL} \ge -\gamma(n,K_0)\frac{\mathcal{D}^2}{V}, \label{chengli}
\end{equation}
where $\gamma\ge 1$ is a constant which only depends on the dimension $n$ and $K_0=K\mathcal{D}^2$. Substituting $G_\text{CL}= G - \frac{\int_\mathcal{M} G}{V}$ into \eqref{chengli} and evaluating the bound at the point where $G=0$, we find that the average of $G$ over the manifold is bounded from above by
\begin{equation}
\frac{\int_\mathcal{M} G}{V} \le \gamma(n,K_0)\frac{\mathcal{D}^2}{V}. \label{chengli2}
\end{equation}
Note that, while warped metrics in string compactifications can have singularities and may not have a bounded Ricci curvature as assumed above (see, e.g., \cite{DeLuca:2021mcj} for a recent discussion), $G$ is defined with respect to the Laplacian of the \emph{smeared} metric for which no singularities appear and the Ricci curvature is bounded. $G$ must therefore satisfy \eqref{chengli2}. Evidently, since $\gamma\ge 1$, our proposal \eqref{est2} is consistent with this requirement. This does of course not exclude the possibility that there might be metrics which violate \eqref{est2}. Nevertheless, it could have been that there is a bound $\frac{\int_\mathcal{M} G}{V} \le \gamma^\prime \frac{\mathcal{D}^2}{V}$ such that $\gamma^\prime \ll \frac{1}{n}$ for some manifolds, which would have been in conflict with \eqref{est2}.
It is reassuring that this is not the case for \eqref{chengli2}, independent of the sign of $K$.

\bigskip

\subsection{Magnitude of the O-plane terms}
\label{sec:b3}

In this section, we argue that the O-planes (and, if present, D-branes) in supercritical dS models\footnote{Our arguments also apply to Minkowski vacua with O-planes.} satisfy the lower bound
\begin{equation}
g_s \sum_i \frac{|T_{p_i}|}{V_i} \gtrsim 1 \label{opd0}
\end{equation}
in the smeared solution, where we ignore $\mathcal{O}(1)$ numerical factors as before and focus on the scaling with $D$.

We first verify this in models without $H_3$ flux (or $H_3$ flux with subleading energy density $|H_3|^2\ll D$).\footnote{In simple models where all sources have the same $p$ and all RR fluxes have the same $q$, we found that setting $H_3=0$, $d=\mathcal{O}(1)$ and $\mathcal{R}_d>0$ yields tachyons in the volume-dilaton moduli subspace. However, there might be more complicated setups avoiding this issue.}
The dilaton and Einstein equations then yield
\begin{equation}
4\pi^2(D-10)= g_s^2\sum_q \frac{2q-D}{4} |F_q|^2 + g_s\sum_i \frac{D-2p_i-4}{4} \frac{T_{p_i}}{2\pi V_i}. \label{dsjdsi}
\end{equation}
Violating the bound \eqref{opd0} by assuming $g_s \sum_i \frac{|T_{p_i}|}{V_i}\ll 1$ would imply $g_s |\sum_i \frac{T_{p_i}}{V_i}|\ll 1$ and $g_s |\sum_i \frac{p_i T_{p_i}}{V_i}|\ll D$ so that the source terms would be subleading in the above equation compared to the term on the left-hand side. Since $q\lesssim D$, it then follows that $g_s^2\sum_q|F_q|^2\gtrsim 1$ or else the equation has no solution. However, this is inconsistent since we know that $g_s |\sum_i \frac{T_{p_i}}{V_i}|$ and $g_s^2\sum_q|F_q|^2$ must be of the same order in $D$ because of \eqref{mn}, \eqref{stab}. We thus conclude that the bound \eqref{opd0} must be satisfied.

A similar bound can also be derived in the asymmetric-orbifold model of \cite{Silverstein:2001xn, Maloney:2002rr}. Recall that $\mathcal{R}_{D-d}=|H_3|^2=0$ in this model and that we should not impose the internal Einstein equations since the internal metric is not dynamical.
However, the dilaton and external Einstein equations (without trace reversal) yield
\begin{equation}
4\pi^2(D-10)= - \frac{d}{4}g_s^2\sum_q  |F_q|^2 - \frac{d+2}{4}g_s\sum_i \frac{T_{p_i}}{2\pi V_i}.
\end{equation}
This can only be satisfied if $g_s \left|\sum_i \frac{T_{p_i}}{V_i}\right| \gtrsim \frac{D}{d}$, which implies \eqref{opd0} and is parametrically stronger unless $d$ is large.

The remaining case are models with leading $|H_3|^2\sim D$. This case is somewhat more involved than the others and we will not present a fully general proof of \eqref{opd0}. We will instead focus on a concrete class of geometries and show that \eqref{opd0} is implied there by tadpole cancellation.
In particular, we consider as the internal space a product of $m$ 3-manifolds, on which we put $H_3$ flux, times possibly another factor without $H_3$ flux:
\begin{equation}
\mathcal{M}_{D-d}=\mathcal{M}_3^{(1)}\times \mathcal{M}_3^{(2)}\times \cdots \times \mathcal{M}_3^{(m)} \times \mathcal{M}_{D-d-3m}. \label{md}
\end{equation}
This is a fairly general class of models as we do not make any further assumptions about the geometries of the various factors.

To derive \eqref{opd0} in these models,
we proceed under the assumption that \eqref{opd0} is violated (i.e., $g_s \sum_i \frac{|T_{p_i}|}{V_i}\ll 1$) and then show that this leads to a contradiction.

We first note that, at leading order in $D$, the Einstein and dilaton equations yield
\begin{equation}
\mathcal{R}_{mn}=\frac{1}{2}|H_3|^2_{mn}, \qquad |H_3|^2= 4\pi^2 (D-10), \label{dsdgfsld}
\end{equation}
where we again used that the RR and source terms have to be of the same order in $D$ and are therefore subleading.
In order to satisfy \eqref{dsdgfsld}, we have to put $H_3$ flux on the 3-cycles in \eqref{md}. We can write $H_3 = \sum_{a=1}^m H_3^{(a)}= \sum_{a=1}^m \frac{h_a \text{dvol}_a}{v_a}$, where $\text{dvol}_a$ and $v_a$ are the volume forms and the volumes of the 3-manifolds and $h_a$ are the quantized fluxes satisfying $\int_{\mathcal{M}_3^{(a)}} H_3^{(a)} = h_a\in\mathbb{Z}$ in string units.
The energy density of the flux on a single 3-cycle is thus
\begin{equation}
|H_3^{(a)}|^2 = \frac{h_a^2}{v_a^2}.
\end{equation}
The corresponding 3-manifold needs to be positively curved and Einstein since
$\mathcal{R}_{mn}=\frac{1}{2}|H_3|^2_{mn} = \frac{h_a^2}{2v_a^2} g_{mn}$ at leading order in $D$. The 3-manifold could thus, e.g., be a 3-sphere or a lens space.
The Bishop-Gromov inequality \cite{bishop1963, bishop1964} now implies that $v_a\le \text{Vol}(S^3)$ for any such manifold, where $\text{Vol}(S^3)$ denotes the volume of the 3-sphere with the same curvature $\mathcal{R}^{(a)}=\frac{3h_a^2}{2v_a^2}$. We thus have $v_a\le \frac{6^{3/2} 2\pi^2}{\mathcal{R}^{(a)3/2}} = \frac{16\pi^2v_a^3}{|h_a|^3} $ and therefore
\begin{equation}
|H_3^{(a)}|^2 \le \frac{16\pi^2}{|h_a|} \label{hk}
\end{equation}
with equality for the 3-sphere.
Since $h_a$ is quantized, this means that $|H_3^{(a)}|^2$ on that particular 3-cycle cannot be larger than $16\pi^2$. However, the equations of motion impose that the total $|H_3|^2$ satisfies $|H_3|^2= 4\pi^2 (D-10)$. We thus conclude that we require $H_3$ flux on at least $m=\frac{D-10}{4}$ 3-cycles and that the energy density on each of these 3-cycles is $|H_3^{(a)}|^2=\mathcal{O}(D^0)$. In addition, we could allow further 3-cycles with small energy densities $|H_3^{(a)}|^2\ll 1$, but these turn out to be irrelevant for the argument so that we will not discuss them further.

Recall now that we require O-planes for dS because of \eqref{mn}. In order to stabilize the dilaton, we then also have to turn on an RR flux $F_q$ according to \eqref{stab}. This generates a flux tadpole $H_3\w F_q$ on $m$ $(q+3)$-cycles $\mathcal{C}_{a}\equiv\mathcal{C}_3^{(a)}\times \mathcal{C}_q$.
For simplicity, we only consider RR flux on a single $q$-cycle here.

The flux tadpole must be cancelled by \mbox{(anti-)}O$p$-planes or \mbox{(anti-)}D$p$-branes
with $q=D-p-4$. We thus have
\begin{equation}
H_3^{(a)} \w F_q = (-1)^{(q+2)(q+3)/2}\frac{Q_{a}}{2\pi V_{a}} \star_a 1 \label{gsfhdhd}
\end{equation}
for all $a$, where $Q_{a}=\sum_{i\in\mathcal{C}_a}Q_{p_i}$ denotes the total charge of all sources contributing to the tadpole on $\mathcal{C}_a$. We further denote by $\star_a1\equiv \star^{(a)}_{q+3}1$ and $V_a$ the corresponding $(q+3)$-dimensional volume forms and volumes, respectively (which should not be confused with the 3-dimensional volumes $v_a$ used earlier).

Crucially, the sources contributing to \eqref{gsfhdhd} wrap $m-1$ of the $m$ $\mathcal{M}_3$ factors in \eqref{md} and thus have $H_3$ on their worldvolumes. We can therefore not have (anti-)D-branes in \eqref{gsfhdhd} due to the Freed-Witten anomaly \cite{Freed:1999vc}.
The sources are thus (anti-)O-planes, which implies that $|T_a| \ge |Q_a|$ and $T_a<0$ on each cycle.\footnote{Here we do not consider exotic positive-tension objects such as O$p^+$-planes, whose consistency conditions would require a separate analysis.}
However, it is not clear to us whether the required intersection pattern of the O-planes can consistently arise from an orientifold involution.\footnote{In particular, note that each O-plane worldvolume shares three dimensions with transverse spaces of other O-planes, which naively means that the worldvolumes cannot be invariant. In familiar examples like type IIA orbifolds with intersecting O6-planes, the corresponding worldvolumes and transverse spaces share an even number of dimensions, at least for the standard orientifold projection \cite[Sec.~4.2]{Danielsson:2011au}.} If this is not the case, (anti-)O-planes are excluded in \eqref{gsfhdhd} as well, which means that the flux tadpole cannot be cancelled in these models. Our earlier assumption $g_s \sum_i \frac{|T_{p_i}|}{V_i}\ll 1$ thus leads to a contradiction and \eqref{opd0} must be satisfied instead, as claimed in the beginning of this section.

On the other hand, if the above intersection pattern does arise in consistent orientifolds, there is no immediate contradiction. However,
the above inequalities for $Q_a$ and $T_a$ then again lead to \eqref{opd0}, as we now show.
In particular, it follows from \eqref{gsfhdhd} that
\begin{equation}
F_{q} = (-1)^{(q+1)(q+2)/2} \frac{\star_a H_3^{(a)}}{|H_3^{(a)}|^2} \frac{Q_{a}}{2\pi V_{a}}.
\end{equation}
We thus obtain
\begin{equation}
g_s^2 |F_{q}|^2 = \frac{g_s^2}{|H_3^{(a)}|^2} \frac{Q_{a}^2}{(2\pi V_{a})^2} \sim g_s^2\frac{Q_{a}^2}{V_{a}^2}, \label{fdzdzzr}
\end{equation}
where in the last step we used that $|H_3^{(a)}|^2=\mathcal{O}(1)$ on each 3-cycle.

Combining our various results, we can now derive the following chain of relations:
\begin{equation}
\left(g_s \sum_a \frac{|T_{a}|}{V_{a}}\right)^2 \ge \left(g_s \sum_a \frac{|Q_{a}|}{V_{a}}\right)^2
\sim D^2 g_s^2 \frac{Q_{a}^2}{V_{a}^2} \sim D^2 g_s^2 |F_{q}|^2 \sim D^2 g_s \left|\sum_{a} \frac{T_{a}}{V_{a}}\right| \sim D^2 g_s \sum_{a} \frac{|T_{a}|}{V_{a}}. \label{ta}
\end{equation}
In the first step, we used that $|T_a| \ge |Q_a|$ on each cycle. In the second and third steps, we used that we have $\mathcal{O}(D)$ 3-cycles and that $\frac{|Q_{a}|}{V_{a}}$ has the same order of magnitude for each of them according to \eqref{fdzdzzr}. In the fourth step, we used the by now familiar fact that the O-plane and RR terms have to be of the same order for dS. Finally, in the last step, we used that $T_a$ on each cycle is negative.
We thus conclude
\begin{equation}
g_s \sum_a \frac{|T_{a}|}{V_{a}} = g_s \sum_i \frac{|T_{p_i}|}{V_i} \gtrsim D^2. \label{ssdgsdg2}
\end{equation}
This is again in contradiction with our earlier assumption $g_s \sum_i \frac{|T_{p_i}|}{V_i} \ll 1$, which confirms our initial claim \eqref{opd0}.\footnote{Aside from the (anti-)O-planes cancelling the flux tadpoles, one may try to put further O-plane/anti-O-plane pairs on other cycles on which no flux tadpole arises. The right-hand side of \eqref{ta} would then read $D^2 g_s^2 |F_q|^2 \sim D^2 g_s |\sum_\text{extra O-planes} \frac{T_{p_i}}{V_{i}} + \sum_{a} \frac{T_{a}}{V_{a}}| \gtrsim D^2 g_s |\sum_{a} \frac{T_{a}}{V_{a}}|$ and thus the same inequality \eqref{ssdgsdg2} would again follow from \eqref{ta}. Another idea is to put brane/anti-brane pairs on cycles without $H_3$ flux such that $D^2 g_s^2 |F_q|^2 \sim D^2 g_s |\sum_\text{extra branes} \frac{T_{p_i}}{V_{i}} + \sum_{a} \frac{T_{a}}{V_{a}}| \ll D^2 g_s |\sum_{a} \frac{T_{a}}{V_{a}}|$. This naively violates \eqref{ssdgsdg2} but introduces open-string instabilities which would have to be stabilized somehow.}

One might wonder whether there is a way to choose the orientation of $F_q$ such that $H_3\w F_q =0$ to avoid generating a flux tadpole and violate \eqref{ssdgsdg2}. However, it is easy to see that this is impossible. For example, consider the case where the 3-manifolds with $H_3$ flux are 3-spheres. We assume that $F_q$ is a harmonic form as usual. Evidently, $F_q$ must then have either three or zero legs on each $S^3$ factor since $b_1=b_2=0$ on an $S^3$. In fact, the only allowed option is \emph{zero} legs since the RR equations of motion impose $H_3\w \star F_q=0$ for harmonic fluxes. It thus follows that $H_3\w F_q$ is non-vanishing and therefore it is unavoidable that a flux tadpole is created. The same argument actually holds more generally if we replace the $S^3$ factors by an arbitrary product of 3-manifolds. Recall that the equations of motion impose a positive Ricci curvature $\mathcal{R}_{mn}=\frac{h_a^2}{2v_a^2}g_{mn}$ on each 3-manifold at leading order in $D$. By a theorem due to Bochner \cite{bochner1946, bochner1948}, it then follows that $b_1=b_2=0$ so that we can again make the same argument leading to $H_3 \w F_q \neq 0$.

It might be possible to avoid the above conclusions by turning on
several RR fluxes or by considering manifolds which are not of the product form \eqref{md}. In particular, preventing a flux tadpole for arbitrary choices of the $h_a$ fluxes would require a positively curved compact manifold with harmonic 3-forms $\omega_{3}^{(a)}$ and a harmonic $q$-form $\omega_{q}$ such that $\omega_{3}^{(a)}\w \omega_{q} =\omega_{3}^{(a)}\w \star\omega_{q}=0$.
Furthermore, this would have to hold for $\gtrsim \mathcal{O}(D)$ independent 3-forms unless the argument around \eqref{hk} (which assumed a product space) does no longer hold. If manifolds with these properties exist, one might be able to fine-tune the O-plane terms to values much smaller than \eqref{opd0}. We anticipate that this may create new problems such as tachyons, but we leave a detailed study of this possibility for future works.

\subsection{Putting everything together}
\label{sec:b4}

We are now ready to put the results of Sections \ref{sec:b1}--\ref{sec:b3} together and state our main argument.
Let us recap what we did so far.
We first argued that backreaction effects are of the order (cf.~\eqref{gb})
\begin{equation}
\mathcal{B}\equiv D g_s\sum_i |T_{p_i}| G_i, \label{jhkozk1}
\end{equation}
where $i$ runs over the various sources and $G_i$ is the Green's function of the Laplacian on the corresponding transverse space.
We then argued that a general estimate for the Green's function is (cf.~\eqref{est2})
\begin{equation}
G_i \gtrsim \frac{\mathcal{D}_i^2}{n_iV_i}. \label{jhkozk2}
\end{equation}
Here $n_i$ is the dimension, $V_i$ is the volume and $\mathcal{D}_i$ is the diameter of the transverse space.
Finally, we argued that the source terms satisfy the bound \eqref{opd0}:
\begin{equation}
g_s\sum_i\frac{|T_{p_i}|}{V_i} \gtrsim 1. \label{tutrwu}
\end{equation}

We now use these results to derive an upper bound on the diameters of the transverse spaces that needs to be satisfied in dS models with a controlled backreaction.
To this end, we substitute \eqref{jhkozk2} in \eqref{jhkozk1} to estimate the backreaction
as
\begin{equation}
\mathcal{B} \gtrsim g_s \sum_{i} \frac{D}{n_i}\frac{|T_{p_i}| \mathcal{D}_i^2}{V_i }. \label{dsgsudhgs}
\end{equation}
Let us further define the weighted average\footnote{Note the slightly different definition compared to $\overline{p}$ in Section \ref{sec:string1}.}
\begin{equation}
\overline{Y} \equiv \frac{\sum_{i} Y_i \frac{|T_{p_i|}}{V_i}}{\sum_{i} \frac{|T_{p_i}|}{V_i}}. \label{avd0}
\end{equation}
Dividing \eqref{dsgsudhgs} by \eqref{tutrwu} and using \eqref{avd0},
it now follows
\begin{equation}
\overline{\frac{\mathcal{D}^2}{n}} \lesssim \frac{\mathcal{B}}{D}. \label{avd1}
\end{equation}
As explained in Section \ref{sec:b1}, a necessary consistency condition is that the backreaction is small on most of the compactification space, i.e.,
\begin{equation}
\mathcal{B} \ll 1,
\end{equation}
since otherwise a singular hole eats up a large fraction of the space and makes the classical approximation unreliable.
We thus have to demand
\begin{equation}
\overline{\frac{\mathcal{D}^2}{n}} \ll \frac{1}{D}, \label{avd}
\end{equation}
i.e., the average diameter of the spaces transverse to the O-planes must be very small at large $D$.
This is one of the main results of this paper.

\bigskip

In many models, \eqref{avd} immediately leads to a contradiction.
In particular, there are classes of manifolds for which rigorous lower bounds on $\mathcal{D}$ are inconsistent with \eqref{avd}. To see this, let us for simplicity compactify on a product space $\mathcal{M}_{D-d}=\mathcal{M}_{D-d-n} \times \mathcal{M}_{n}$ where $\mathcal{M}_{D-d-n}$ is wrapped by the O-planes and $\mathcal{M}_{n}$ is the transverse space. Now assume, for example, that $\mathcal{M}_{n}$ is a K\"ahler-Einstein manifold with Ricci curvature $\mathcal{R}_{mn}= (n-1)Kg_{mn}$ for some $K>0$. 
Then, as shown in \cite{Collins:2022nux}, the diameter is bounded from below as $\mathcal{D}\gtrsim (nK)^{-1/2}$ up to $\mathcal{O}(1)$ factors.
We can now determine $K$ using the Einstein and dilaton equations. For $\mathcal{R}_d\ge 0$, the equations imply that the scalar curvature of $\mathcal{M}_{n}$ satisfies
\begin{align}
\mathcal{R}_n &\le
\frac{3}{2} 4\pi^2(D-10) - \frac{n-3}{4} g_s \sum_i \frac{T_{p_i}}{2\pi V_i}
< \frac{3}{2} 4\pi^2(D-10) + \frac{n}{4} g_s \sum_i \frac{|T_{p_i}|}{2\pi V_i}, \label{gkgskjgfj}
\end{align}
where the second inequality holds because $-\sum_i \frac{T_{p_i}}{V_i} =|\sum_i \frac{T_{p_i}}{V_i}|\le \sum_i \frac{|T_{p_i}|}{V_i}$ for dS. The last term on the right-hand side is bounded by \eqref{tutrwu}.
Let us assume for the moment that \eqref{tutrwu} is saturated, i.e., $g_s \sum_i\frac{|T_{p_i}|}{V_i} \sim 1$. Since $n\lesssim D$, \eqref{gkgskjgfj} then yields $\mathcal{R}_n\lesssim D$ and therefore $K\lesssim \frac{D}{n^2}$ up to $\mathcal{O}(1)$ factors. The results of \cite{Collins:2022nux} thus imply that the diameter of $\mathcal{M}_{n}$ is bounded by $\mathcal{D}^2 \gtrsim \frac{n}{D}$, which is inconsistent with \eqref{avd}.

The same conclusion actually follows if \eqref{tutrwu} is far from saturation, e.g., for $g_s\sum_i\frac{|T_{p_i}|}{V_i} \sim D$.
Substituting this into \eqref{gkgskjgfj} yields $\mathcal{R}_n\lesssim n D$ and therefore $K\lesssim \frac{D}{n}$ so that the lower bound on the diameter is relaxed to $\mathcal{D}^2 \gtrsim \frac{1}{D}$. One might thus naively think that, for large enough $n$, there is no contradiction with the upper bound \eqref{avd} anymore. However, this is too premature. Indeed, going again through our derivation of \eqref{avd} but this time using $g_s\sum_i\frac{|T_{p_i}|}{V_i}  \sim D$, we find that now there is a stronger upper bound $\overline{\frac{\mathcal{D}^2}{n}} \ll \frac{1}{D^2}$. We thus again arrive at a contradiction as before.

We conclude that
K\"ahler-Einstein manifolds with $K>0$ are incompatible with small backreaction. This is purely a consequence of the fact that the geometry does not allow small enough diameters.
It was conjectured in \cite{Collins:2022nux} that similar lower bounds on $\mathcal{D}$ hold more generally for any Einstein manifold with $K>0$, even when they are not K\"ahler. For example, this is certainly true if the transverse space is a (product of) sphere(s). If the conjecture of \cite{Collins:2022nux} holds with the same bound $\mathcal{D}\gtrsim (nK)^{-1/2}$ that holds for K\"ahler-Einstein manifolds, this is again in tension with our upper bound \eqref{avd} and implies that the O-plane backreaction must be large in such models.

A possible concern with the above discussion might be
that compactifications with localized sources can be singular and do not have a bounded Ricci curvature so that the results of \cite{Collins:2022nux} do not apply. However, note that all objects that appear above are defined with respect to the smeared metric, which \emph{does} have a bounded Ricci curvature (with $K>0$ in the above examples).
As emphasized before in Section \ref{sec:b1}, our viewpoint is that the classical description of a compactification with O-planes only has a chance of being reliable if we are allowed to work with the smeared solution.
In particular, we expect that the smeared solution becomes classically exact in limits where the strong-backreaction regions near the O-planes shrink to points (right-hand side of Fig.~\ref{smeared}). What we showed above is that it is not self-consistent to approach such limits on some manifolds, which indicates that the classical description breaks down (left-hand side of Fig.~\ref{smeared}).

\bigskip

A lower bound on $\mathcal{D}$ contradicting \eqref{avd} can also arise from flux quantization.
As a simple example, consider the case where $\mathcal{M}_n$ is an $n$-torus with radius $R$. Since the diameter is $\mathcal{D} \sim\sqrt{n}R$, \eqref{avd} states that the O-plane backreaction is large in this model unless the radius is parametrically small: $R^2 \ll \frac{1}{D}$.
Now we put $H_3$ flux somewhere on the torus. The energy density is $|H_3|^2= \frac{\sum_a h_a^2}{(2\pi R)^6} \ge \frac{1}{(2\pi R)^6}$, where $h_a$ are the quantized fluxes through the various 3-cycles.
For $\mathcal{R}_d\ge 0$, the Einstein and dilaton equations imply
\begin{equation}
|H_3|^2 \le 4\pi^2(D-10) - \frac{n-1}{2}g_s \sum_i \frac{T_{p_i}}{2\pi V_i} <
4\pi^2(D-10) + \frac{n}{2}g_s \sum_i\frac{|T_{p_i}|}{2\pi V_i}.
\end{equation}
Assuming saturation of the bound \eqref{tutrwu}, this yields $|H_3|^2 \lesssim D$. At the same time, we have $|H_3|^2\gtrsim R^{-6}$ so that
\begin{equation}
\frac{\mathcal{D}^2}{n} \sim R^2 \gtrsim \frac{1}{D^{1/3}}, \label{rgkr}
\end{equation}
which is inconsistent with \eqref{avd}. As before, one can furthermore verify that the same conclusion follows if \eqref{tutrwu} is far from saturation. The lower bound \eqref{rgkr} then becomes weaker but at the same time the upper bound \eqref{avd} becomes stronger so that there is again a contradiction.

To summarize, the above discussion shows that large classes of models cannot satisfy \eqref{avd}, either for purely geometric reasons or due to the physical requirement of flux quantization.
In such models, the classical description is not self-consistent and we should not trust any dS solutions obtained this way.

\bigskip

Another interesting example violating \eqref{avd} is the asymmetric-orbifold model studied in \cite{Silverstein:2001xn, Maloney:2002rr}.
The model has O-planes and anti-O-planes of codimension $n=D-d$.\footnote{There are also spacetime-filling (anti-)O-planes with $n=0$, which are, however, not important for the argument.} The transverse space $\mathcal{M}_n$ in this case is a torus which is fixed by the orbifolding at the self-dual radius $R\sim 1$. We thus have $\mathcal{D}\sim \sqrt{n}$, which apparently is in conflict with \eqref{avd}.
Since the internal space is not geometric in this model, it is a priori not obvious whether the violation of \eqref{avd} can be interpreted as a large O-plane backreaction and to what extent this is pathological. However, we consider it plausible that one should still impose \eqref{avd}, as we now explain. The idea of an asymmetric orbifold is to orbifold by a T-duality symmetry \cite{Narain:1986qm}.
Since T-duality requires an appropriate isometry \cite{Buscher:1987sk, Rocek:1991ps, Hassan:1999mm}, we expect that the orbifold picture is reliable in a regime where the O-planes in the compactification can be treated as approximately smeared. Otherwise their backreaction breaks the isometries and there is no T-duality by which we can orbifold to begin with. This suggests that we should again consider \eqref{jhkozk1} as a proxy for backreaction and demand that it is small just like in the geometric models. Since we have already shown that \eqref{jhkozk2} and \eqref{tutrwu} hold in the asymmetric-orbifold model (see Sections \ref{sec:b2} and \ref{sec:b3}), it then follows that \eqref{avd} would have to hold for consistency, as discussed at the beginning of this section.

\subsection{Backreaction and the KK scale}
\label{sec:b5}

Aside from the model-dependent obstructions we just discussed, there is another problem with the bound \eqref{avd}. Indeed, the fact that the diameters need to be  small in string units raises a concern about the validity of the classical effective action, as small diameters are typically associated with a breakdown of such an approach.
In particular, making the compact space small is expected to increase the curvature and the flux densities and thus have the effect that higher-derivative corrections are no longer suppressed (cf.~Section \ref{sec:string}). At the same time, a small compact space implies a high KK scale so that it may no longer make sense to keep the KK modes dynamical in the effective field theory while integrating out the string states.
We thus expect a tension between the three different consistency conditions of small backreaction, small $\alpha^\prime$ corrections and a scale separation between the KK and string scales.

We can make this intuition more precise. Let us consider an $n$-dimensional compact Riemannian manifold without boundary and with bounded Ricci curvature such that $\mathcal{R}_{mn}u^mu^n\ge (n-1) K$ for some $K\in\mathbb{R}$ and all unit tangent vectors $u^m$.
As stated before, warped compactifications with localized sources may be singular and need not have a bounded Ricci curvature but the quantities of interest in this section (diameters, volumes, etc.) are defined with respect to the smeared metric, which is smooth and does satisfy such a bound. As explained in Section \ref{sec:b1},
large deviations from the predictions of the smeared solution are expected to signal a breakdown of the effective field theory so that it makes sense to demand that this is not the case.

Under the above conditions, we can estimate the first non-zero eigenvalue $\lambda_1$ of the Laplacian in terms of $\mathcal{D}$, $n$ and $K$. Historically, such estimates were often obtained assuming in addition $K> 0$ or $K= 0$ (see, e.g., \cite{Lichnerowicz1958, li1980, zhong1984}) but there are also bounds which are valid for $K<0$.
For our purpose, a useful bound applying to any $K\in\mathbb{R}$ is due to Shi and Zhang \cite{shi2007}:
\begin{equation}
\lambda_1 \ge \frac{\pi^2}{\mathcal{D}^2}+ \frac{n-1}{2}K. \label{sz}
\end{equation}
Similar earlier results with numerically slightly weaker bounds were obtained in \cite{cai1991, chen1997}.

Interestingly, \eqref{sz} provides a model-independent \emph{lower} bound on the diameter in terms of the curvature and the KK scale, which is in potential conflict with our \emph{upper} bound \eqref{avd} related to the backreaction.\footnote{Note that, depending on the metric, \eqref{sz} can strongly underestimate the true diameter. For example, for the round sphere, we have $\mathcal{D}=\pi R$ but using $\lambda_1=nR^{-2}$, $K=R^{-2}$ in \eqref{sz} only gives the estimate $\mathcal{D} \ge \frac{\sqrt{2}\pi R}{\sqrt{n+1}}$, which is parametrically weaker at large $n$. The true tension with the upper bound \eqref{avd} can therefore be parametrically stronger than indicated by \eqref{sz}.}
To make this manifest, we first average \eqref{sz} as defined in \eqref{avd0} over the transverse spaces associated to the various localized sources. We thus find
\begin{equation}
\overline{n\lambda_1} - \frac{\overline{n(n-1) K}}{2} \gtrsim \overline{\frac{n}{\mathcal{D}^2}}.
\end{equation}
Using $\overline{\frac{n}{\mathcal{D}^2}}\ge \left(\overline{\frac{\mathcal{D}^2}{n}}\right)^{-1}$\footnote{This follows from \eqref{avd0} and the fact that $(\sum_i x_i y_i)(\sum_i \frac{1}{x_i}y_i)\ge (\sum_i y_i)^2$ for $x_i,y_i>0$.} and substituting \eqref{avd1} then yields
\begin{equation}
\overline{n\lambda_1} - \frac{\overline{n(n-1) K}}{2} \gtrsim \frac{D}{\mathcal{B}}. \label{main}
\end{equation}
As anticipated, \eqref{main} directly relates three crucial consistency conditions for the validity of the $D$-dimensional spacetime effective action: small $\alpha^\prime$ corrections (controlled by $K$), small O-plane backreaction (controlled by $\mathcal{B}$), and a separation between the KK scale and the string scale (controlled by $\lambda_1$). This confirms and makes quantitative our claim of a tension between these three consistency conditions in supercritical dS models.

In particular, \eqref{main} states that a small backreaction $\mathcal{B} \ll 1$ requires at least some of the manifolds transverse to the O-planes to satisfy either
\begin{equation}
\lambda_1 \gg \frac{D}{n} \qquad \text{or} \qquad |K| \gg \frac{D}{n(n-1)},\quad K<0. \label{goihfhdhf}
\end{equation}
The second possibility is actually excluded by the Einstein equations, which yield
\begin{equation}
\mathcal{R}_{mn} = \frac{1}{2}|H_3|^2_{mn}+\frac{1}{2}g_s^2 \sum_q |F_q|^2_{mn} +\frac{1}{d}g_{mn}\mathcal{R}_d +\frac{1}{2} g_s\sum_i \frac{T_{p_i}}{2\pi V_i} \left( g_{mn} - \Pi^{(i)}_{mn}\right).
\end{equation}
For dS, this implies
\begin{equation}
\mathcal{R}_{mn}u^mu^n > - \frac{1}{2}g_s\sum_i \frac{|T_{p_i}|}{2\pi V_i}
\end{equation}
for all unit tangent vectors $u^m$. Since $K$ is defined as the minimum of the Ricci curvature, $\mathcal{R}_{mn}u^mu^n\ge -(n-1) |K|$, it follows
\begin{equation}
|K| = \frac{1}{2(n-1)}g_s\sum_i \frac{|T_{p_i}|}{2\pi V_i},
\end{equation}
where the right-hand side is bounded by \eqref{tutrwu}. If \eqref{tutrwu} is saturated, we have $|K| \sim \frac{1}{n-1}$. Since $n\lesssim D$, it follows that the second possibility in \eqref{goihfhdhf} is ruled out.
We thus conclude that a small backreaction is tied to a large KK scale (in string units):
\begin{equation}
\lambda_1 \gg \frac{D}{n}. \label{lam}
\end{equation}
We will not discuss in detail the case where \eqref{tutrwu} is far from saturation (e.g., $g_s\sum_i \frac{|T_{p_i}|}{V_i}\sim D$) but just note here that the bound \eqref{lam} then becomes even stronger. To see this, one has to rederive \eqref{avd1} and \eqref{goihfhdhf} using $g_s\sum_i \frac{|T_{p_i}|}{V_i}\sim D$, which introduces an extra factor of $D$ in these inequalities and, as a consequence, in \eqref{lam}, leading to an even larger KK scale $\lambda_1 \gg \frac{D^2}{n}$.

The problem with \eqref{lam} is the following.
We know that the KK scale has to be smaller than the squared masses of the first excited string states, $\lambda_1 \ll M_s^2$. Otherwise it would not make sense to consider a $D$-dimensional spacetime effective action where string states are integrated out but KK modes are dynamical.
Hence, $M_s^2$ needs to be large in the dS background:
\begin{equation}
M^2_s \gg \lambda_1 \gg \frac{D}{n}. \label{ms}
\end{equation}

This certainly looks suspicious. Recall that quantizing the $D=10$ superstring in flat space yields $M_s^2\sim 1$ \cite{Polchinski:1998rr}. For general $D$, the squared masses of the lowest string excitations scale linearly with $D$ and $D$ appears with a minus sign (see, e.g., \cite[Secs.~1.3, 1.4]{Polchinski:1998rq} for the bosonic string). However, this result should be taken with caution as flat space is not a consistent background for $D\neq 10$. In a non-trivial background with curvature and fluxes, the squared masses presumably receive further $\mathcal{O}(D)$ contributions (recall from Section \ref{sec:string} that the equations of motion impose $\gtrsim \mathcal{O}(D)$ curvature and energy densities). This suggests $M_s^2\lesssim D$, although we are not aware of explicit results deriving this.
Assuming that $M_s^2\sim 1$, \eqref{ms} is violated, which indicates that the effective field theory breaks down, either due to large O-plane backreaction or light string states below the KK scale. If there are backgrounds with $M_s^2\sim D$, the tension is relaxed to some extent and we only observe an immediate problem in models with small $n$. Of course, in that case, there may nevertheless be a conflict with bounds on the geometry or with flux quantization, as discussed in Section \ref{sec:b4}.

\bigskip

It is instructive to compare the results of this section with (AdS) flux compactifications in the \emph{critical} string theories. We consider again the DGKT vacua. These solutions admit a limit of large 4-form flux, where backreaction and $\alpha^\prime$ corrections become small and the KK modes become light in string units \cite{DeWolfe:2005uu, Camara:2005dc, Junghans:2020acz, Marchesano:2020qvg, Cribiori:2021djm}.
The key difference to supercritical flux vacua is that, in the critical case, there is no dilaton tadpole $\sim D-10$ in the equations of motion, which has the consequence that the source terms need not satisfy \eqref{tutrwu}. Instead, they scale like $g_s\sum_i \frac{|T_{p_i}|}{V_i} \sim N^{-3/2}$ in DGKT, where $N\gg 1$ is an unbounded discrete parameter related to the 4-form fluxes. Repeating the derivation of \eqref{main} under this new assumption for the source terms yields
\begin{equation}
\overline{n\lambda_1} - \overline{\frac{n(n-1)K}{2}} \gtrsim \frac{D}{N^{3/2} \mathcal{B}}.
\end{equation}
We can simplify this using that the Ricci curvature is zero and that the backreaction is of the order $\mathcal{B}=1/N$ in DGKT \cite{Junghans:2020acz, Marchesano:2020qvg}. Using further $D=10$, $n=3$, we arrive at
\begin{equation}
\overline{\lambda_1} \gtrsim N^{-1/2}.
\end{equation}
At large $N$, the bound on the KK scale thus goes to zero in string units and no problem arises for the effective field theory.
This is in agreement with estimating the KK scale as $\lambda_1\sim \mathcal{V}^{-1/3}$, where $\mathcal{V}\sim N^{3/2}$ is the Calabi-Yau volume.\footnote{Note that estimating the KK scale in terms of the volume yields the correct parametric dependence on $N$ since the large-$N$ limit is a uniform rescaling of the internal manifold and does not create any parametrically large anisotropies.}
This provides a non-trivial cross-check of our formula \eqref{main} and the various arguments leading up to it.

\section{Discussion - de Sitter and the Dine-Seiberg problem}
\label{sec:discussion}

In the preceding sections, we argued that supercritical dS models are not under control at large $D$ since the required O-planes generate large backreaction corrections that invalidate the classical description. Let us now widen the scope and discuss the possible implications of this result on a more general level.

Although our arguments are not a rigorous proof, they show a generic control problem with supercritical dS which appears to be difficult, if not impossible, to avoid.
As we will review momentarily, this is in line with a number of other recent results showing similar problems in various dS scenarios in string theory. Interestingly, these scenarios are technically very different from one another but the problems found there appear to be universal. In particular, a recurring observation is that string theory does not admit dS vacua in a perturbative regime where
the scalar potential is dominated by a few leading terms and an infinite number of string and backreaction corrections is \emph{parametrically} controlled (i.e., their suppression is guaranteed by a small parameter and we do not have to make ad hoc assumptions about unknown $\mathcal{O}(1)$ coefficients).
This is fundamentally different from AdS where a parametric control appears to be possible. In particular, in DGKT, all known corrections are suppressed by inverse powers of the 4-form fluxes, which can be chosen arbitrarily large \cite{DeWolfe:2005uu, Camara:2005dc, Junghans:2020acz, Marchesano:2020qvg, Cribiori:2021djm}.

We now briefly review the control problems discovered in the various dS scenarios:
\begin{itemize}
\item \emph{KKLT scenario} \cite{Kachru:2003aw}:
Many string and backreaction corrections to the KKLT effective field theory
are suppressed by powers of the Calabi-Yau volume $\mathcal{V}$,
where $\mathcal{V}^{2/3}\sim \ln (|W_0|^{-1})$ in terms of the flux superpotential $W_0$.
Since the IIB flux landscape is finite \cite{Douglas:2003um, Ashok:2003gk, Denef:2004ze, Grimm:2020cda, Bakker:2021uqw}, there is a minimal $|W_0|$ in a given compactification and therefore a maximal volume. Hence, unlike DGKT, there is necessarily a limit to how strongly we can suppress the corrections.
Nevertheless, one might hope that the minimal $W_0$ is small enough so that in practice it serves as a control parameter and hence KKLT vacua exist in a perturbative regime.
See \cite{Lust:2022lfc, Demirtas:2021nlu} for a recent debate about whether this is possible for AdS vacua.
Here we are concerned with the dS case for which there is by now strong evidence that perturbative control is in fact \emph{not} possible.
Assuming an anti-brane uplift, the volume in the putative dS vacuum is constrained by the requirement that the AdS vacuum energy is balanced with the uplift term, which leads to a number of problems \cite{Carta:2019rhx, Gao:2020xqh, Blumenhagen:2022dbo}.\footnote{
See also \cite{DallAgata:2022abm, Kallosh:2022fsc} for an independent debate about a possible instability of the anti-brane uplift based on goldstino condensation.} In particular, it was shown in \cite{Gao:2020xqh} that the volume is too small to prevent a large backreaction of the uplift on the bulk geometry, causing a large part of the latter to become formally singular (``singular-bulk problem'').
It is plausible that string theory resolves these singularities by replacing them with a strongly curved stringy geometry \cite{Carta:2021lqg}. However, as emphasized in \cite{Gao:2020xqh}, the real problem is not the singularity itself but the fact that we lose control over the K\"ahler potential, which is only known in the supergravity regime \cite{Grimm:2004uq}. The singular-bulk problem thus implies that the KKLT effective field theory is unreliable in the dS case since string corrections are not under control. Moreover, \cite{Gao:2020xqh} also discussed a number of generalizations of the original KKLT scenario, including models with large $h^{1,1}$, and argued that they are also unlikely to avoid the singular-bulk problem.

\item \emph{LARGE-volume scenario (LVS)} \cite{Balasubramanian:2005zx}: A similar control problem occurs for LVS dS vacua with anti-brane uplift.
The volume scales like $\mathcal{V} \sim \e^{\mathcal{O}(1)/g_s}$ in this scenario, where $g_s$ is a function of the flux parameters and the $\mathcal{O}(1)$ number in the exponent depends on the topology of the Calabi-Yau 3-fold on which one compactifies. As in the KKLT scenario, $g_s$ is bounded since the IIB flux landscape is finite. Nevertheless, a moderately small $g_s$ is sufficient to make the volume exponentially large, which naively suggests an excellent control over string and backreaction corrections. However, it turns out that this is not the case \cite{Junghans:2022exo, Gao:2022fdi, Junghans:2022kxg, Hebecker:2022zme, Schreyer:2022len, ValeixoBento:2023nbv}. Indeed, one finds that
there is always at least one type of correction which is not suppressed by any small parameter, making the naive LVS effective field theory unreliable \cite{Junghans:2022exo, Junghans:2022kxg}. Similar results were obtained in \cite{Hebecker:2022zme, Schreyer:2022len} taking into account the polarization channel of the anti-D3-brane(s) into an NS5-brane and in \cite{ValeixoBento:2023nbv} for a less strongly warped variant of the LVS.\footnote{\cite{Hebecker:2022zme, Schreyer:2022len} also proposed to avoid these problems by considering a modified version of KKLT/LVS where the anti-D3/NS5-brane sits in a throat with \emph{large} curvature corrections. It is not obvious how to reliably derive the effective field theory describing this regime. In any case, the proposal of \cite{Hebecker:2022zme, Schreyer:2022len} would be consistent with our above claim that string theory does not admit dS in a \emph{perturbative} regime where string corrections aside from a small number of leading terms are negligible.}
A possible way out would be
if the numerical coefficients of the various corrections could be tuned small, e.g., through their complex-structure-moduli dependence \cite{Junghans:2022exo, Junghans:2022kxg}. However, it is currently not known how to compute these coefficients and whether it is possible to make several of them small simultaneously. An exception are curvature corrections to the anti-D3/NS5-brane, which have been computed and shown to \emph{not} have small coefficients \cite{Junghans:2022exo, Hebecker:2022zme, Schreyer:2022len}.

\item \emph{classical dS scenario} \cite{Silverstein:2007ac, Hertzberg:2007wc}: Another popular scenario proposes to search for dS vacua of the (critical) type II string theories in the classical regime. This means that, in contrast to KKLT and LVS, one attempts to stabilize all moduli classically using fluxes, while all (non-)perturbative string corrections are assumed to be negligible.
While there is no rigorous proof that suppressing string corrections is impossible in this setting (see, e.g., \cite{Andriot:2019wrs} for a discussion), several no-go results strongly support this conclusion if one demands parametric control.
In particular, \cite{Junghans:2018gdb, Banlaki:2018ayh} argued that IIA compactifications with O6-planes do not admit dS vacua in the asymptotic large-volume/small-coupling regime. This is true in setups with arbitrary fluxes and Ricci curvature and also includes non-trivial scaling limits where multiple fields become large \cite[Sec.~3.2]{Junghans:2018gdb}. The same arguments can also be applied to compactifications with O4 or O5-planes \cite{Andriot:2019wrs}. For compactifications with O8-planes, it was shown in \cite{Cribiori:2019clo} that they do not admit dS vacua in a regime where string corrections have a negligible effect on the cosmological constant (see also the discussion in \cite[Sec.~5.2]{Junghans:2020acz}).\footnote{Recall that, although string corrections locally blow up near the O-planes, their integrated effect in the lower-dimensional effective field theory would have to be small in order to claim a consistent solution (cf.~Section \ref{sec:shc}).} All of these setups therefore at best admit dS vacua in regimes where the classical approximation is inconsistent without further knowledge about the string corrections.

\end{itemize}

The universal problems observed with these different dS scenarios strongly suggest an underlying principle.
There is a special case where there is indeed a simple explanation for this, namely the case where the scalar potential has no large parameters (such as flux numbers or the dimension $D$). The absence of dS (and other) vacua in the perturbative regime is then a consequence of the well-known Dine-Seiberg problem \cite{Dine:1985he}. The key insight is that, in string theory, perturbative control is linked to moving towards the asymptotics of the moduli space. Consider, for example, the potential $V(\phi)$ for the dilaton $\phi$, which controls the string coupling via $\e^\phi=g_s$. At small coupling $g_s \ll 1$, we can expand $V$ in powers of $g_s$. In the absence of other large parameters, it is then clear that $V$ must be dominated by the leading term $V \sim \e^{-2\phi}$, implying a runaway unless $V=0$. Hence, it immediately follows that (A)dS minima with respect to the dilaton can only exist in the strong-coupling region $g_s\sim 1$ where the $g_s$-expansion breaks down and we lose perturbative control. An analogous argument can also be made for the volume modulus, which controls the $\alpha^\prime$ expansion \cite{Denef:2008wq}. Furthermore, a generalized version of the Dine-Seiberg problem arises for dS in \emph{any} asymptotic region of the moduli space if one assumes the swampland distance conjecture \cite{Ooguri:2018wrx}. Hence, if the potential has no large parameters, the Dine-Seiberg problem and its generalization in \cite{Ooguri:2018wrx} provide a simple explanation for why string theory does not have dS vacua in regimes of perturbative control.

Interestingly, the results reviewed above now suggest that dS is also forbidden in the more general case where the potential contains large parameters like fluxes or the dimension $D$, although naively one should not have a (generalized) Dine-Seiberg problem there (see \cite[Sec.~2]{Junghans:2018gdb} for a discussion).
In particular, as the DGKT AdS vacua exemplify, a potential with unbounded parameters can have minima at arbitrarily large field values.
It is therefore a priori not clear why the same should not work for dS vacua.
A possible explanation is that there exists an analogue of the Dine-Seiberg problem
in the asymptotics of the discrete-parameter space spanned by the fluxes and the dimension.
To see whether this is true or not, it may be useful to consider a generalized notion of distance for discrete parameters, perhaps along the lines of \cite{Shiu:2022oti} and \cite{Bonnefoy:2020uef, DeBiasio:2021yoe, Lust:2021mgj}. We hope to come back to these ideas in future work.

\section{Conclusions}
\label{sec:concl}

Flux compactifications of supercritical string theories are an interesting arena to test and challenge our prejudices from the lamppost of the critical theories.
In this paper, we analyzed whether supercritical dS models allow a better control over dangerous string corrections than critical dS models like KKLT, LVS or the classical dS scenario.
One might have hoped that the new parameter $D$ that arises in the supercritical case can be used as a control parameter in the sense that string corrections are parametrically suppressed at large $D$.
However, we argued that this is not the case and that supercritical dS models instead have control problems much like their cousins in the critical theories.
The key problem we identified is that the O-plane backreaction cannot be made small in the supercritical models.
This has the effect of generating singular holes which eat up large parts of the classical spacetime and thus make the classical description unreliable.
Supercritical dS models thus look like the picture on the left in Fig. \ref{smeared}, whereas a controlled model would look like the picture on the right.

In particular, we argued that in order to control the backreaction effects one would have to impose an upper bound on the average diameters of the transverse spaces of the localized sources.
On the other hand, \emph{lower} bounds on such diameters are satisfied due to geometric constraints or physical requirements like flux quantization. A lower bound on diameters also needs to be imposed in order to avoid light string states below the KK scale.
We found that the upper bound from controlling the backreaction and the lower bounds are often in tension with each other, ruling out reliable dS constructions in the corresponding models.

As byproducts of our analysis, we derived a number of technical results that are useful beyond the specific applications in this paper. Most notably, we computed the leading backreaction corrections to the smeared solution in a general flux compactification from $D$ to $d$ dimensions for an arbitrary distribution of O-planes and D-branes. We also argued for a general estimate for Green's functions on compact manifolds (and therefore for the backreaction corrections) in terms of their diameter and volume which remains valid for metrics with large anisotropies and in any dimension.

Although we believe that the arguments presented in this paper are compelling,
they are in part based on examples so that
we cannot exclude that models exist which somehow evade our constraints. It would be very interesting to look for such counter-examples or rule them out, for example, through a rigorous proof of the proposed bound on Green's functions in Section \ref{sec:b2} or by studying the tadpole argument in Section \ref{sec:b3} for a more general class of manifolds. These are interesting research problems where geometry and string theory intersect.

In view of some of our results, it would also be important to learn more about supercritical string theory beyond the classical approximation, for example, concerning the large-$D$ scaling of the tower of massive string states or of higher-derivative corrections to the classical effective action.
Knowing the scaling of the string states would in particular allow to make the constraints of Sections \ref{sec:string3} and \ref{sec:b5} related to KK/string-scale separation more precise. The scaling of the higher-derivative terms would be interesting because of the results of Section \ref{sec:string1}, where we showed that the curvature or some energy densities of the NSNS/RR field strengths are always large in the string frame. While the majority of this paper focussed on dS vacua, this particular result applies to AdS, Minkowski and dS vacua alike, and we argued that it may be in tension with controlling the higher-derivative terms. However, as pointed out in \cite{Harribey:2018xvs, DeLuca:2021pej, Flauger:2022hie}, such a conclusion may be avoided if the coefficients of the higher-derivative terms are sufficiently strongly suppressed at large $D$. Computing these coefficients would therefore be very interesting.

Our focus throughout this work was on parametric control over string and backreaction corrections at large $D$, which is one of the key features of the supercritical-dS scenario proposed in \cite{Silverstein:2001xn, Maloney:2002rr, Dodelson:2013iba, Harribey:2018xvs, DeLuca:2021pej, Flauger:2022hie}. Our results do therefore not exclude the weaker claim of numerical control for some finite $D>10$. Although it is unclear why one should expect a simultaneous numerical suppression of \emph{all} corrections in the absence of a common scaling factor, it would in principle be very interesting to explore this possibility further. However, numerical control is in practice extremely hard to verify and therefore a much less attractive scenario since it would require to compute all relevant higher-derivative corrections and the full non-linear O-plane backreaction including all numerical coefficients.

We finally note that many arguments in this paper involved a non-trivial interplay of various constraints from geometry (e.g., estimates for volumes, diameters and Green's functions) and string theory (e.g., flux quantization, tadpole cancellation, absence of tachyons and anomalies). It is quite intriguing that all these different constraints needed to work together in order to uncover the control problems of naively consistent dS solutions.
Which of the constraints are relevant turned out to depend on the specific classes of models considered.
For example, the arguments in Section \ref{sec:b3} were rather different for models with or without $H_3$ flux but led to the same conclusion. As discussed in Section \ref{sec:discussion}, there is furthermore strong evidence in the recent literature that the popular dS scenarios in the critical theories share similar control problems, even though they are technically quite different from one another and from the supercritical models. This universality suggests an underlying structure in string theory. Understanding this structure and learning something from it about the nature of dark energy will be a crucial task for the future.

\section*{Acknowledgments}

I would like to thank Eva Silverstein for a useful correspondence. This research was funded by the Austrian Science Fund (FWF) under project number P 34562-N.

\appendix

\section{Minkowski solutions}
\label{app:mink}

We present here a simple class of Minkowski solutions in supercritical string theories. These are analogous to the no-scale solutions of type IIB but do not require any O-planes. We consider a spacetime of the form
\begin{align}
\mathbb{R}^{1,d-1} \times  &\underbrace{S^3 \times \cdots \times S^3}  \times \mathbb{T}^{D-d-3m} \notag \\[-0.5em]
& \qquad\! \scriptstyle{m \text{ times}}
\end{align}
with $H_3$ flux on each of the 3-spheres. More generally, one could replace the 3-spheres by other positively curved Einstein 3-manifolds and the torus by any Ricci-flat manifold.

We can write $H_3 = \sum_{a=1}^m \frac{h_a \text{dvol}_a}{v_a}$ with $h_a\in\mathbb{Z}$, where $v_a$ are the 3-sphere volumes.
The equations of motion then fix the radii $R_a$ of the spheres to $R_a^2 = \frac{|h_a|}{4\pi^2}$ and impose an additional condition
\begin{equation}
\sum_{a=1}^m \frac{16\pi^2}{|h_a|}=4\pi^2 (D-10) \label{nsm}
\end{equation}
on the fluxes. One can furthermore verify that the dilaton and the torus radii are flat directions at the classical level.

The condition \eqref{nsm} is satisfied, e.g., for the choice $m=\frac{D-10}{4}$ and $h_a=1$ with $10-d+\frac{D-10}{4}$ unstabilized torus radii. An example with the dilaton as the only flat direction is $m=\frac{D-d}{3}$, $d=10$, $D=10 + 16\cdot 9k$, $k\in \mathbb{N} $ and $h_{a}=h_{a+1}=1$, $h_{a+2}=4$ for $a=1,4,7,\ldots$.

We emphasize that these are solutions to the equations of motion of the \emph{classical} $D$-dimensional spacetime effective action. Whether these solutions correspond, exactly or approximately, to genuine vacua of a supercritical string theory is not obvious due to the possibly large $\alpha^\prime$ corrections (see Section \ref{sec:string}). Assuming that the large-$D$ suppression of the latter is strong enough that the solutions are reliable, it would be interesting to try to stabilize the dilaton using non-perturbative effects and/or loop corrections. The former case is a supercritical analogue of the KKLT scenario but without O-planes and the associated tadpole restrictions and backreaction effects, which are at the heart of various control problems of the original scenario.

\section{Comments on non-linear O-plane backreaction}
\label{app:caveat}

In this appendix, we elaborate on a caveat related to the discussion in Section \ref{sec:val}.
In particular, we argued there that the regime $g_sG \gtrsim 1/D$ where the O-planes backreact non-linearly is intimately linked to large string corrections and thus a breakdown of the classical solution.
While this is true in all examples we are aware of, we do not know precisely what happens at $g_sG \gtrsim 1/D$ unless we actually compute the full solution non-perturbatively in $g_sG$ and identify the relevant (perturbative and non-perturbative) string corrections that resolve the putative classical singularity.
We can therefore not rule out the possibility that, in some backgrounds, the backreaction at $g_sG \gtrsim 1/D$ modifies the classical solution less violently than expected.

Let us discuss this a bit more explicitly. We consider again a single source with tension $\frac{T_p}{2\pi}=-1$ and ask what happens as we approach the source. In particular, let us consider a scenario where we have three qualitatively different regimes: a) a regime far away from the source where $g_sG\ll 1/D$ such that the backreaction is a small perturbation, b) an intermediate regime $1/D \lesssim g_sG\ll C(D)$ in which the backreaction is already non-linear but string corrections are still negligible, and c) the near-source regime $g_sG \gtrsim C(D)$ where string corrections become relevant and resolve the singularities generated by the classical backreaction. The condition for small string corrections is thus $g_sG\ll C$, which for large enough $C$ is much weaker than the condition $g_sG\ll 1/D$ for small backreaction.

We do not know whether such ``mildly backreacting'' orientifold backgrounds exist in string theory. However, we now show in a simple toy example that this may require rather strong assumptions. In particular, we find in our model that the backreaction is not mild in the above sense unless the solution remains classical up to energy densities which are \emph{exponentially} large in $D$ (in string units).

To see this, let us assume that the non-linear backreaction yields
\begin{equation}
\e^{2\phi}|F_{n-3}|^2 = \frac{g_s^2|\tilde F_{n-3}|^2}{(1-g_sG)^{(D-4)/4}}. \label{nlb}
\end{equation}
Note that this agrees with the first term in \eqref{fn0} at linear order in $g_sG$. For $D=10$, $n=6$ it furthermore reduces to \eqref{fgkp}, which is the correct all-order expression in the GKP solution \cite{Giddings:2001yu} (see also \cite{Blaback:2010sj} for T-dual versions with a similar behavior).\footnote{Other choices such as $\e^{2\phi}|F_{n-3}|^2 = \frac{g_s^2|\tilde F_{n-3}|^2}{(1-\frac{D-4}{6}g_sG)^{3/2}}$ are also consistent with \eqref{fn0}, \eqref{fgkp} but yield a finite-distance singularity at $g_sG\gtrsim 1/D$ and thus do not realize the mild scenario described above.} We stress again, however, that this is a toy model. In general, the non-linear backreaction will not lead to such a simple expression, but it suffices to illustrate our point.

We now assume that the smeared solution scales like $g_s^2|\tilde F_{n-3}|^2\sim D^x$
and that string corrections become relevant for $\e^{2\phi}|F_{n-3}|^2\gtrsim D^y$ for some $x,y\in \mathbb{R}$. We further take $y>x$ since otherwise the smeared solution is out of control.
Using this in \eqref{nlb}, we conclude that string corrections are important for
\begin{equation}
g_sG \gtrsim 1- D^{\frac{4(x-y)}{D-4}} \sim \frac{4(y-x)\ln(D)}{D} \label{largesc2}
\end{equation}
at large $D$.
The regime where the classical solution breaks down thus agrees parametrically (up to a logarithmic factor) with the regime $g_sG\gtrsim 1/D$ where the backreaction becomes non-linear. This is true regardless of the precise energy scale at which the string corrections become relevant, as long as it is power-law in $D$.

On the other hand, if the solution remains classical up to \emph{exponentially} large $\e^{2\phi}|F_{n-3}|^2\gtrsim \e^{yD}$, \eqref{nlb} yields
\begin{equation}
g_sG \gtrsim 1-\e^{-4y} \sim D^0
\end{equation}
at large $D$. In contrast to the power-law case, the regime of large string corrections now differs parametrically from the regime $g_sG\gtrsim 1/D$ of large backreaction.
Our toy example thus illustrates that the connection between non-linear backreaction and string corrections is to some extent sensitive to the nature of the latter.\footnote{But much less so than the arguments in Section \ref{sec:string}. Indeed, the problem with suppressing string corrections observed there is already resolved if the solution remains classical for $\mathcal{O}(D)$ energy densities, which is power-law rather than exponential in $D$.}

\bibliographystyle{utphys}
\bibliography{groups}

\end{document}